\documentclass[twocolumn]{aastex631}
\usepackage{longtable}
\usepackage{soul}

\begin{document}

\title{SN~2020udy: A new piece of the homogeneous bright group in the diverse Iax subclass}

\correspondingauthor{Mridweeka Singh}
\email{mridweeka.singh@iiap.res.in, yashasvi04@gmail.com}

\author[0000-0001-6706-2749]{Mridweeka Singh}
\affiliation{Indian Institute of Astrophysics, Koramangala 2nd Block, Bangalore 560034, India}

\author[0000-0002-6688-0800]{Devendra K. Sahu}
\affiliation{Indian Institute of Astrophysics, Koramangala 2nd Block, Bangalore 560034, India}

\author[0000-0003-4769-4794]{Barnab\'as Barna}
\affiliation{Department of Experimental Physics, Institute of Physics, University of Szeged, H-6720 Szeged, D{\'o}m t{\'e}r 9, Hungary}

\author[0000-0002-3884-5637]{Anjasha Gangopadhyay}
\affiliation{Hiroshima Astrophysical Science Center, Hiroshima University, Higashi-Hiroshima, Japan}

\author[0000-0001-6191-7160]{Raya Dastidar}
\affiliation{Millennium Institute of Astrophysics (MAS), Nuncio Monsenor Sòtero Sanz 100, Providencia, Santiago RM, Chile}
\affiliation{Instituto de Astrofísica, Universidad Andres Bello, Fernandez Concha 700, Las Condes, Santiago RM, Chile}

\author[0000-0002-0525-0872]{Rishabh Singh Teja}
\affiliation{Indian Institute of Astrophysics, Koramangala 2nd Block, Bangalore 560034, India}
\affiliation{Pondicherry University, R.V. Nagar, Kalapet, 605014, Puducherry, India}

\author[0000-0003-1637-267X]{Kuntal Misra}
\affiliation{Aryabhatta Research Institute of observational sciencES, Manora Peak, Nainital, 263001, India}

\author[0000-0003-4253-656X]{D. Andrew Howell}
\affiliation{Las Cumbres Observatory, 6740 Cortona Drive Suite 102, Goleta, CA, 93117-5575, USA}
\affiliation{Department of Physics, University of California, Santa Barbara, CA 93106-9530, USA}

\author[0000-0002-7334-2357]{Xiaofeng Wang}
\affiliation{Physics Department, Tsinghua University, Beijing, 100084, China}
\affiliation{Beijing Planetarium, Beijing Academy of Science and Technology, Beijing, 100044, China}

\author{Jun Mo}
\affiliation{Physics Department, Tsinghua University, Beijing, 100084, China}

\author{Shengyu Yan}
\affiliation{Physics Department, Tsinghua University, Beijing, 100084, China}

\author[0000-0002-1125-9187]{Daichi Hiramatsu}
\affiliation{Center for Astrophysics \textbar{} Harvard \& Smithsonian, 60 Garden Street, Cambridge, MA 02138-1516, USA}
\affiliation{The NSF AI Institute for Artificial Intelligence and Fundamental Interactions, USA}
\affiliation{Las Cumbres Observatory, 6740 Cortona Drive Suite 102, Goleta, CA, 93117-5575, USA}
\affiliation{Department of Physics, University of California, Santa Barbara, CA 93106-9530, USA}

\author[0000-0002-7472-1279]{Craig Pellegrino}
\affiliation{Las Cumbres Observatory, 6740 Cortona Drive Suite 102, Goleta, CA, 93117-5575, USA}
\affiliation{Department of Physics, University of California, Santa Barbara, CA 93106-9530, USA}

\author[0000-0003-3533-7183]{G. C. Anupama}
\affiliation{Indian Institute of Astrophysics, Koramangala 2nd Block, Bangalore 560034, India}

\author[0000-0001-9275-0287]{Arti Joshi}
\affiliation{Institute of Astrophysics, Pontificia Universidad Católica de Chile, Av. Vicuña MacKenna 4860, 7820436, Santiago, Chile}

\author[0000-0002-4924-444X]{K. Azalee Bostroem}
\affiliation{DiRAC Institute, Department of Astronomy, University of Washington, Box 351580, U.W., Seattle, WA 98195, USA}

\author[0000-0002-7472-1279]{Jamison Burke}
\affiliation{Las Cumbres Observatory, 6740 Cortona Drive Suite 102, Goleta, CA, 93117-5575, USA}
\affiliation{Department of Physics, University of California, Santa Barbara, CA 93106-9530, USA}

\author[0000-0001-5807-7893]{Curtis McCully}
\affiliation{Las Cumbres Observatory, 6740 Cortona Drive Suite 102, Goleta, CA, 93117-5575, USA}
\affiliation{Department of Physics, University of California, Santa Barbara, CA 93106-9530, USA}

\author{Rama Subramanian V}
\affiliation{Department of Sciences, Amrita School of physical sciences, Amrita Vishwa Vidyapeetham, Coimbatore, Tamil Nadu, 641112}

\author{Gaici Li}
\affiliation{Physics Department, Tsinghua University, Beijing, 100084, China}

\author{Gaobo Xi}
\affiliation{Physics Department, Tsinghua University, Beijing, 100084, China}

\author[0000-0001-5879-8762]{Xin Li}
\affiliation{Beijing Planetarium, Beijing Academy of Science and Technology, Beijing, 100044, China}

\author{Zhitong Li}
\affiliation{National Astronomical Observatories of China, Chinese Academy of Sciences, Beijing, 100012, China}

\author[0000-0003-4524-6883]{Shubham Srivastav}
\affiliation{Astrophysics Research Centre, School of Mathematics and Physics, Queen’s University Belfast, Belfast BT7 1NN, UK}

\author[0000-0002-5742-8476]{Hyobin Im}
\affiliation{Korea Astronomy and Space Science Institute, 776 Daedeokdae-ro, Yuseong-gu, Daejeon 34055, Republic of Korea}
\affiliation{Korea University of Science and Technology (UST), 217 Gajeong-ro, Yuseong-gu, Daejeon 34113, Republic of Korea}

\author[0000-0002-7708-3831]{Anirban Dutta }
\affiliation{Indian Institute of Astrophysics, Koramangala 2nd Block, Bangalore 560034, India}
\affiliation{Pondicherry University, Chinna Kalapet, Kalapet, Puducherry 605014, India}



\begin{abstract}

We present optical observations and analysis of a bright type Iax SN~2020udy hosted by NGC 0812. The light curve evolution of SN~2020udy is similar to other bright Iax SNe. Analytical modeling of the quasi bolometric light curves of SN~2020udy suggests that 0.08$\pm$0.01 M$_{\odot}$ of $^{56}$Ni would have been synthesized during the explosion. Spectral features of SN~2020udy are similar to the bright members of type Iax class showing weak Si {\sc II} line. The late-time spectral sequence is mostly dominated by Iron Group Elements (IGEs) with broad emission lines. Abundance tomography modeling of the spectral time series of SN~2020udy using TARDIS indicates stratification in the outer ejecta, however, to confirm this, spectral modeling at a very early phase is required. After maximum light, uniform mixing of chemical elements is sufficient to explain the spectral evolution. Unlike the case of normal type Ia SNe, the photospheric approximation remains robust until +100 days, requiring an additional continuum source. Overall, the observational features of SN~2020udy are consistent with the deflagration of a Carbon-Oxygen white dwarf.

\end{abstract}

\keywords{Supernovae (1668) --- Type Ia supernovae (1728)}

\section{Introduction} \label{sec:intro}

Thermonuclear supernovae (SNe), also known as type Ia SNe, are the outcome of the explosive burning of Carbon-Oxygen (CO) white dwarfs. They are known as one parameter family and are extensively used in cosmology as standard candles \citep{1993ApJ...413L.105P,1999AJ....118.1766P}.  With the increasing sample, diversity has been observed among type Ia SNe leading to their subclassification \citep{2017hsn..book..317T}. Distinguishing different subclasses can help improve the precision of their distance measurements \citep{2009ApJ...699L.139W} and identify their physical origins \citep{2013Sci...340..170W}. 

There are different subtypes of type Ia SNe having some similarities and dissimilarities. Amongst them, type Iax SNe \citep{2003PASP..115..453L,2006AJ....131..527J,2013ApJ...767...57F} are one of the peculiar subclasses of SNe Ia, having low luminosity (M$_{r}$ = $-$12.7 mag, \citealt{Karambelkar_2021} to M$_{V}$ = $-$18.4 mag, \citealt{2011ApJ...731L..11N}) and lower energy budget \citep[][and references therein]{2017hsn..book..375J}. The light curves of SNe Iax are characterized by a faster rise to the maximum and post-maximum decline in the bluer bands than normal type Ia SNe \citep{2016A&A...589A..89M,2017A&A...601A..62M,2017hsn..book..375J,2018MNRAS.478.4575L}. Due to the lack of fairly good pre-maximum coverage for a good fraction of type Iax SNe, precise measurement of peak absolute magnitudes and rise time is difficult. This also poses a problem in putting strong observational constraints on the explosion models. The occurrence rate of SNe Iax is $\sim$ 5\% to 30\% of that of normal type Ia SNe \citep{2013ApJ...767...57F,Miller_2017,2022MNRAS.511.2708S}. 
 
As a class, type Iax SNe show homogeneous spectral evolution \citep{2006AJ....132..189J,2017hsn..book..375J} with low expansion velocities at maximum ranging from 2000 to 8000\,km s$^{-1}$ \citep{2009AJ....138..376F,2014A&A...561A.146S}. The early spectra of SNe Iax show Fe {\sc III} features along with features due to intermediate mass elements (IMEs), similar to 1991T-like SNe. During the late phase, type Iax SNe exhibit significant differences from the type Ia class. Strong and wide P-Cygni lines dominate the optical wavelengths until 4-6 months after explosion. Even later, type Iax SNe do not enter into a fully nebular phase. Permitted spectral lines mainly of Fe co-exist with forbidden emission lines in the late time spectra of all type Iax SNe \citep{2014ApJ...786..134M,2015A&A...573A...2S,2016MNRAS.461..433F}. The spectral synthesis of SN 2014dt (the only example with continuous observations from its maximum to +550 days) showed that the assumption of an expanding photosphere provides a remarkable match with the observed spectral evolution during the first $\sim100$ days. At even later epochs, the approximation is capable of reproducing the P-Cygni lines formed by Fe, Ca, and Na \citep{2023ApJ...951...67C}. 

The progenitor system and explosion scenario of type Iax SNe has been a matter of debate for many years. Being low luminosity and less energetic events, type Iax SNe hint towards a different progenitor scenario from type Ia SNe. High-resolution pre-explosion images of a few type Iax SNe, obtained with the HST, are used to identify the progenitor systems of these events. SN\,2012Z is one such type Iax SNe, hosted by a nearby galaxy NGC 1309. \cite{2014Natur.512...54M} analyzed the pre-explosion images of SN\,2012Z and suggested that a binary consisting of a white dwarf and Helium star is one of the most plausible progenitor systems for SN\,2012Z.
Another type Iax with pre-explosion images is SN\,2014dt, and a similar progenitor system has been suggested as one of the possibilities \citep{2015ApJ...798L..37F} for this SN. However, except for a few type Iax SNe \citep{2013ApJ...767...57F,2023Natur.615..605G},  helium is not detected in spectroscopic studies \citep{2015ApJ...799...52W,2019MNRAS.487.2538J,2019A&A...622A.102M}.  

The observed explosion parameters of many of the bright type Iax SNe are successfully explained by the pure deflagration of a CO white dwarf \citep{2012ApJ...761L..23J,2013MNRAS.429.2287K,2014MNRAS.438.1762F}. However, several other explosion models have been proposed to explain the observational properties of bright type Iax SNe such as pulsational delayed detonation (PDD, \citealt{Baron_2012,2014MNRAS.441..532D}), deflagration to detonation transition (DDT, \citealt{2013MNRAS.429.1156S,2013MNRAS.436..333S}), etc. The deflagration of CO white dwarf cannot explain the observables of faint members of this class. Instead, they can be explained by the deflagration of a hybrid Carbon–Oxygen–Neon (CONe) white dwarf \citep{2014ApJ...789L..45M,2015MNRAS.450.3045K}.

SN~2020udy was spotted by \citet{2020TNSTR2929....1N} using automated detection software AMPEL \citep{2019A&A...631A.147N} in association with Zwicky Transient Facility (ZTF, \citealt{2019PASP..131f8003B, Fremling_2020}) on 24 September 2020 and was classified as type Iax SN (\citealt{2020TNSAN.185....1N,2020TNSCR2940....1N}). The SN exploded in a spiral galaxy NGC 0812 at a redshift of 0.017222 \citep{1999PASP..111..438F}. The SN was located at R.A.(J2000.0)  02$^h$06$^m$49.35$^s$, Dec.(J2000.0) 44$^o$35$'$15.29${''}$, 52$^{''}$.82 N and 23$^{''}$.13 W from the center of the host galaxy. 

\cite{2023MNRAS.525.1210M} have presented an analysis of SN~2020udy. The very early detection of the SN allowed them to place strict limits on companion interaction. They ruled out the possibility of a main sequence star with mass 2 and 6 M$_{\odot}$, to be the companion, however a helium star with a narrow range of viewing angle is suggested as a probable companion. They have shown that the light curve and spectra of SN~2020udy are in good agreement with the deflagration model of a CO white dwarf, specifically the N5-def model of \citet{2014MNRAS.438.1762F}.

This paper presents a detailed photometric and spectroscopic analysis of SN~2020udy. The light curve has been modeled using analytical prescription proposed by \citet{1982ApJ...253..785A} and \citet{2008MNRAS.383.1485V}. To confirm the line identification and expansion velocity of the ejecta, the early spectral sequence is modeled with {\texttt{SYNAPPS}. One-dimensional radiative transfer code TARDIS has been used to perform abundance tomography modeling of the entire observed spectral sequence. Section \ref{observation and data reduction} gives details about observations and methods used to reduce the data of SN~2020udy. In Section 3 we estimate the distance, the explosion time, and the line-of-sight extinction of the SN. Section 4 presents the photometric properties and the modeling of the pseudo-bolometric light curve of SN~2020udy. Spectroscopic features of SN~2020udy, the evolution of the photospheric velocity, and spectral modeling are presented in Section \ref{spectral evolution}. A brief discussion of the observational features of SN~2020udy and their comparison with a few explosion scenarios proposed for the type Iax class are given in Section \ref{discussion}. Section \ref{summary} summarizes our results.    

\section{Observations and data reduction}
\label{observation and data reduction}

Optical photometric follow-up of SN 2020udy began $\sim$ 5 days after discovery and continued up to 130 days, with the 1 m Las Cumbres Observatory (LCO) telescopes \citep{2013PASP..125.1031B} under the Global Supernova Project and 80 cm  Tsinghua-NAOC Telescope (TNT, \citealt{2008ApJ...675..626W,2012RAA....12.1585H}, National Astronomical Observatories of China). The observations were carried out in the {\it BgVri} photometric bands. 

The LCO photometric data were reduced using the lcogtsnpipe routines \citep{2016MNRAS.459.3939V}, which performs point-spread-function (PSF) photometry of the stars and the SN. The instrumental {\it BV} magnitudes were calibrated to the Vega system using the APASS catalog \citep{APASS} \footnote{\url{https://www.aavso.org/aavso-photometric-all-sky-survey-data-release-1}} and the instrumental {\it gri} magnitudes were calibrated to the AB system using the SDSS catalog \citep{Gunn_2006}.

The pre-processing of the photometric data, obtained with the 80 cm TNT was carried out following standard procedures using a custom Fortran program. The photometry was then performed by the automatic pipeline ZrutyPhot (Jun Mo, et al., in prep.). 
The pipeline utilizes the Software for Calibrating AstroMetry and Photometry (SCAMP, \citealt{2006ASPC..351..112B}) and the IRAF\footnote{Image Reduction and Analysis Facility, {\url{http://iraf.noao.edu/}}} Daophot package. The TNT instrumental $BV$ magnitudes of the SN were calibrated to the Vega system and the instrumental $gri$ magnitudes were calibrated to the AB system using the PanSTARRS (Panoramic Survey Telescope and Rapid Response System\footnote{\url{https://catalogs.mast.stsci.edu/panstarrs}}) catalog. The optical photometry of SN~2020udy are tabulated in Table~\ref{tab:photometric_observational_log_2020udy}. The errors mentioned in Table~\ref{tab:photometric_observational_log_2020udy} are obtained by propagating the photometric and the calibration errors in quadrature.

\begin{longtable*}{*{9}{p{\dimexpr(\textwidth-20\tabcolsep)/9\relax}}}
\caption{Optical photometric measurements of SN 2020udy}
\\
\hline
Date    &   JD$^\dagger$   &   Phase$^\ddagger$ 	&  ~~B       				& ~~g          	 	& ~~ V                		& ~~r   	        	 & ~~i      & Telescope$\star$             \\
        &                  &   (Days)           	& (mag)     				& (mag)            		 & (mag)             		&(mag) 			&(mag)      &      \\
\hline  
\endfirsthead

\multicolumn{9}{c}%
    {{\bfseries \tablename\ \thetable{} -- Continued from previous page}} \\
    \hline
    Date    &   JD$^\dagger$   &   Phase$^\ddagger$ 	&   ~~B       		&      ~~g          	 	&  ~~V                		& ~~r   	    & ~~i      & Telescope$\star$             \\
    \hline
    \endhead

\hline 
\multicolumn{9}{l}{\textsuperscript{} $^\dagger$ JD 2,459,000+ ,
$^\ddagger$ Phase has been calculated with respect to $B$$_{max}$ = 2459130.53,
$^\star$ 1: LCO, 2: TNT
} \\
\multicolumn{9}{r}{\textit{Continued on next page}} \\ 
\hline
    \endfoot
    
    \hline
    \endlastfoot

2020-09-29   &  122.24  &  -8.29    &    17.27$\pm$0.02	           &  17.16$\pm$0.01    &   17.27$\pm$0.02     & 17.17$\pm$0.01  &   17.34$\pm$0.02      &       2    \\       		
2020-10-01   &  124.23  &  -6.30    &    16.91$\pm$0.03	           &  16.82$\pm$0.01    &   16.92$\pm$0.02     & 16.85$\pm$0.01  &   17.02$\pm$0.01      &       2    \\    
2020-10-02   &  124.98  &  -5.55    &    16.80$\pm$0.04            &  16.78$\pm$0.05    &   16.70$\pm$0.03     & 16.72$\pm$0.03  &   16.94$\pm$0.07      &       1    \\    
2020-10-04   &  127.19  &  -3.34    &    16.80$\pm$0.02	           &  16.58$\pm$0.01    &   16.66$\pm$0.02     & 16.57$\pm$0.00  &   16.69$\pm$0.01      &       2    \\    
2020-10-05   &  127.98  &  -2.55    &    16.56$\pm$0.03		        &  16.61$\pm$0.04    &   16.42$\pm$0.03     & 16.34$\pm$0.04  &   16.49$\pm$0.05      &       1    \\    
2020-10-05   &  128.23  &  -2.30    &    16.74$\pm$0.01	           &  16.55$\pm$0.00    &   16.57$\pm$0.01     & 16.48$\pm$0.01  &   16.61$\pm$0.01      &       2    \\    
2020-10-07   &  130.18  &  -0.35    &    16.72$\pm$0.01	           &  16.46$\pm$0.00    &   16.41$\pm$0.01     & 16.35$\pm$0.00  &   16.47$\pm$0.01      &       2    \\    
2020-10-08   &  130.88  &  0.35     &    16.49$\pm$0.09		       &  16.46$\pm$0.13    &   16.10$\pm$0.16     & 16.28$\pm$0.04  &        &       1    \\    
2020-10-11   &  133.73  &  3.20     &    16.74$\pm$0.05		      &  16.55$\pm$0.06    &   16.23$\pm$0.11     & 16.18$\pm$0.03  &   16.30$\pm$0.04     &       1    \\    
2020-10-16   &  139.23  &  8.70     &    17.38$\pm$0.01	           &  16.86$\pm$0.00    &   16.45$\pm$0.00     & 16.21$\pm$0.00  &   16.25$\pm$0.00     &       2    \\    
2020-10-17   &  139.63  &  9.10     &    17.39$\pm$0.03		   &  16.99$\pm$0.03    &   16.45$\pm$0.02     & 16.20$\pm$0.04  &   16.19$\pm$0.03     &       1    \\    
2020-10-17   &  140.22  &  9.69     &    17.47$\pm$0.02	           &  16.93$\pm$0.01    &   16.48$\pm$0.01     & 16.23$\pm$0.01  &   16.22$\pm$0.01     &       2    \\    
2020-10-18   &  141.19  &  10.66    &    17.63$\pm$0.01	           &  17.04$\pm$0.01    &   16.53$\pm$0.01     & 16.26$\pm$0.00  &   16.24$\pm$0.01     &       2    \\    
2020-10-19   &  142.23  &  11.70    &    17.74$\pm$0.02	           &  17.13$\pm$0.01    &   16.59$\pm$0.01     & 16.28$\pm$0.00  &   16.24$\pm$0.01     &       2    \\    
2020-10-20   &  142.77  &  12.24    &    17.77$\pm$0.09	           &  17.40$\pm$0.05    &   16.70$\pm$0.09     & 16.32$\pm$0.05  &   16.22$\pm$0.06     &       1    \\    
2020-10-21   &  144.21  &  13.68    &    18.03$\pm$0.02	           &  17.38$\pm$0.01    &   16.75$\pm$0.01     & 16.39$\pm$0.00  &   16.31$\pm$0.00     &       2    \\    
2020-10-22   &  144.94  &  14.41    &    18.06$\pm$0.04		   &  17.68$\pm$0.06    &   16.84$\pm$0.05     & 16.50$\pm$0.10  &   16.33$\pm$0.08     &       1    \\    
2020-10-22   &  144.94  &  14.41    &    18.10$\pm$0.07		   &  17.61$\pm$0.05    &   16.86$\pm$0.06     & 16.51$\pm$0.09  &   16.31$\pm$0.09     &       1    \\    
2020-10-23   &  146.33  &  15.80    &    18.29$\pm$0.02	           &  17.59$\pm$0.00    &   16.92$\pm$0.01     & 16.50$\pm$0.00  &   16.40$\pm$0.00     &       2    \\    
2020-10-25   &  148.29  &  17.76    &    18.51$\pm$0.03	           &  17.79$\pm$0.01    &   17.05$\pm$0.01     & 16.61$\pm$0.01  &   16.48$\pm$0.01     &       2    \\    
2020-10-26   &  149.28  &  18.75    &    18.52$\pm$0.03	           &  17.79$\pm$0.01    &   17.12$\pm$0.01     & 16.67$\pm$0.00  &   16.49$\pm$0.01     &       2    \\    
2020-10-31   &  153.90  &  23.37    &    18.98$\pm$0.10		   &  18.43$\pm$0.06    &   17.38$\pm$0.03     & 16.77$\pm$0.05  &   16.56$\pm$0.05     &       1    \\    
2020-10-31   &  153.91  &  23.38    &    19.03$\pm$0.09		   &  18.38$\pm$0.06    &   17.48$\pm$0.03     & 16.70$\pm$0.06  &        &       1    \\    
2020-11-03   &  157.28  &  26.75    &    18.96$\pm$0.11	           &  18.34$\pm$0.05    &   17.65$\pm$0.04     & 17.18$\pm$0.01  &   16.98$\pm$0.02     &       2    \\    
2020-11-04   &  157.75  &  27.22    &    19.12$\pm$0.06		   &  18.40$\pm$0.06    &   17.64$\pm$0.03     & 17.03$\pm$0.05  &   16.76$\pm$0.05     &       1    \\             
2020-11-04   &  157.75  &  27.22    &    19.01$\pm$0.05		   &  18.40$\pm$0.04    &   17.62$\pm$0.04     & 17.03$\pm$0.06  &   16.72$\pm$0.06     &       1    \\             
2020-11-04   &  158.23  &  27.70    &    19.19$\pm$0.11		   &  18.36$\pm$0.04    &   17.68$\pm$0.03     & 17.14$\pm$0.01  &   16.96$\pm$0.01     &       2    \\             
2020-11-06   &  160.23  &  29.70    &    19.23$\pm$0.09		   &  18.47$\pm$0.02    &   17.75$\pm$0.02     & 17.27$\pm$0.01  &   17.10$\pm$0.01     &       2    \\             
2020-11-07   &  161.22  &  30.69    &    18.94$\pm$0.08		   &  18.38$\pm$0.03    &   17.62$\pm$0.03     & 17.25$\pm$0.01  &   17.03$\pm$0.02     &       2    \\             
2020-11-08   &  161.87  &  31.34    &    19.08$\pm$0.13		   &  18.52$\pm$0.04    &   17.61$\pm$0.02     & 17.19$\pm$0.04  &   17.14$\pm$0.01     &       1    \\             
2020-11-08   &  162.23  &  31.70    &    19.17$\pm$0.05		   &  18.50$\pm$0.03    &   17.62$\pm$0.02     & 17.21$\pm$0.04  &       &       2    \\             
2020-11-08   &  162.24  &  31.71    &           &      &   17.79$\pm$0.02     & 17.32$\pm$0.01  &      &       2    \\             
2020-11-09   &  163.26  &  32.73    &    19.32$\pm$0.04		   &  18.55$\pm$0.01    &   17.85$\pm$0.01     & 17.40$\pm$0.01  &   17.21$\pm$0.01     &       2    \\             
2020-11-10   &  163.89  &  33.36    &    19.22$\pm$0.03		   &  18.63$\pm$0.04    &   17.79$\pm$0.01     & 17.33$\pm$0.05  &   17.19$\pm$0.06     &       1    \\             
2020-11-10   &  163.89  &  33.36    &    19.27$\pm$0.04		   &  18.65$\pm$0.04    &   17.77$\pm$0.02     &  &   17.09$\pm$0.05     &       1    \\            
2020-11-13   &  167.26  &  36.73    &    19.40$\pm$0.09		   &      &   17.92$\pm$0.02     &  &        &       2    \\             
2020-11-14   &  167.75  &  37.22    &    19.35$\pm$0.05		   &  18.75$\pm$0.04    &   17.99$\pm$0.04     &   &   17.37$\pm$0.09     &       1    \\             
2020-11-14   &  167.76  &  37.23    &    19.34$\pm$0.05		   &  18.77$\pm$0.04    &   17.98$\pm$0.05     &   &        &       1    \\             
2020-11-15   &  169.21  &  38.68    &    19.34$\pm$0.05		   &  18.63$\pm$0.02    &   17.97$\pm$0.02     & 17.55$\pm$0.01  &   17.40$\pm$0.01     &       2    \\             
2020-11-18   &  171.76  &  41.23    &    19.22$\pm$0.12       &     &   17.83$\pm$0.08     &   &        &       1    \\             
2020-11-18   &  171.77  &  41.24    &    19.24$\pm$0.08       &      &   17.89$\pm$0.08     &   &        &       1    \\             
2020-11-22   &  175.84  &  45.31    &           &  18.70$\pm$0.06    &       &   &   17.61$\pm$0.05      &       1    \\    
2020-11-22   &  175.84  &  45.31    &          &  18.71$\pm$0.06    &       &   &   17.71$\pm$0.09      &       1    \\    
2020-11-26   &  179.81  &  49.28    &    19.44$\pm$0.07     &  18.74$\pm$0.08    &   17.95$\pm$0.05     & 17.93$\pm$0.13  &   17.72$\pm$0.06      &       1    \\           
2020-11-26   &  179.82  &  49.29    &          &     &   17.87$\pm$0.05     &   &   17.89$\pm$0.06      &       1    \\            
2020-11-27   &  181.08  &  50.55    &    19.44$\pm$0.26		   &  18.62$\pm$0.08    &   18.27$\pm$0.07     & 17.88$\pm$0.03  &   17.73$\pm$0.03      &       2    \\           
2020-11-28   &  182.05  &  51.52    &    19.32$\pm$0.17		   &      &   18.23$\pm$0.06     & 17.91$\pm$0.03  &   17.75$\pm$0.03      &       2    \\           
2020-11-29   &  183.02  &  52.49    &    19.54$\pm$0.21	           &  18.72$\pm$0.09    &   18.19$\pm$0.06     & 17.92$\pm$0.04  &   17.80$\pm$0.03      &       2    \\           
2020-11-30   &  184.14  &  53.61    &    19.46$\pm$0.17	           &  18.82$\pm$0.07    &   18.18$\pm$0.07     & 17.95$\pm$0.03  &   17.78$\pm$0.03      &       2    \\           
2020-12-02   &  185.71  &  55.18    &    19.53$\pm$0.09		   &  18.99$\pm$0.06    &   18.22$\pm$0.05     & 17.96$\pm$0.06  &   17.82$\pm$0.08      &       1    \\    
2020-12-02   &  185.72  &  55.19    &    19.52$\pm$0.09		   &  18.91$\pm$0.06    &   18.29$\pm$0.04     & 17.91$\pm$0.06  &   17.89$\pm$0.09      &       1    \\    
2020-12-02   &  186.07  &  55.54    &    19.92$\pm$0.30	           &  18.89$\pm$0.11    &   18.10$\pm$0.07     & 17.93$\pm$0.04  &   17.87$\pm$0.04      &       2    \\    
2020-12-03   &  187.09  &  56.56    &    19.48$\pm$0.14	           &  18.82$\pm$0.05    &   18.32$\pm$0.04     & 18.03$\pm$0.02  &   17.92$\pm$0.02      &       2    \\    
2020-12-04   &  188.09  &  57.56    &    19.57$\pm$0.09	           &  18.86$\pm$0.04    &   18.28$\pm$0.02     & 17.98$\pm$0.01  &   17.89$\pm$0.02      &       2    \\    
2020-12-08   &  191.71  &  61.18    &    19.49$\pm$0.15		   &     &     &   &         &       1    \\    
2020-12-09   &  193.02  &  62.49    &    19.55$\pm$0.05     	   &  18.90$\pm$0.02    &   18.38$\pm$0.02     & 18.12$\pm$0.01  &   18.02$\pm$0.01      &       2    \\    
2020-12-10   &  194.05  &  63.52    &    19.64$\pm$0.05	           &  18.94$\pm$0.02    &   18.45$\pm$0.02     & 18.18$\pm$0.01  &   18.04$\pm$0.01      &       2    \\    
2020-12-11   &  195.05  &  64.52    &    19.73$\pm$0.07	           &  18.95$\pm$0.02    &   18.45$\pm$0.02     & 18.17$\pm$0.01  &   18.07$\pm$0.02      &       2    \\    
2020-12-12   &  195.70  &  65.17    &    19.71$\pm$0.04		   &      &   18.42$\pm$0.10     &   &        &       1    \\    
2020-12-12   &  195.71  &  65.18    &    19.69$\pm$0.03		   &         &   18.44$\pm$0.10     &   &         &       1    \\    
2020-12-16   &  200.11  &  69.58    &    19.68$\pm$0.04	           &  18.99$\pm$0.02    &   18.53$\pm$0.02     & 18.29$\pm$0.01  &   18.20$\pm$0.01      &       2    \\    
2020-12-17   &  201.09  &  70.56    &    19.53$\pm$0.06	           &  18.98$\pm$0.03    &   18.51$\pm$0.02     & 18.26$\pm$0.02  &   18.17$\pm$0.02      &       2    \\    
2020-12-18   &  201.76  &  71.23    &    19.61$\pm$0.04            &  18.87$\pm$0.06    &   18.44$\pm$0.02     & 18.24$\pm$0.08  &   18.10$\pm$0.06      &       1    \\    
2020-12-18   &  201.78  &  71.25    &    19.61$\pm$0.04            &  18.91$\pm$0.05    &   18.47$\pm$0.02     & 18.15$\pm$0.08  &        &       1    \\    
2020-12-25   &  208.71  &  78.18    &    19.79$\pm$0.07     	   &  18.97$\pm$0.10    &   18.35$\pm$0.05     &  &         &       1    \\    
2020-12-25   &  208.71  &  78.18    &    19.84$\pm$0.07     	   &  18.97$\pm$0.11    &   18.32$\pm$0.05     &  &        &       1    \\    
2021-01-02   &  216.72  &  86.19    &    19.96$\pm$0.08     	   &  19.24$\pm$0.11    &   18.85$\pm$0.04     & 18.61$\pm$0.09  &         &       1    \\    
2021-01-02   &  216.72  &  86.19    &    19.93$\pm$0.08     	   &  19.27$\pm$0.11    &   18.84$\pm$0.05     &  &       &       1    \\    
2021-01-08   &  222.70  &  92.17    &    19.90$\pm$0.06     	   &    &   18.64$\pm$0.04     &   &        &       1    \\    
2021-01-08   &  222.71  &  92.18    &    19.99$\pm$0.06     	   &     &   18.69$\pm$0.04     &  &      &       1    \\    
2021-01-12   &  227.03  &  96.50    &              &  19.25$\pm$0.14    &   18.89$\pm$0.19     & 18.73$\pm$0.08  &   18.53$\pm$0.07      &       2    \\     
2021-01-14   &  228.71  &  98.18    &                &      &        & 18.83$\pm$0.05  &         &       2    \\     
2021-01-14   &  228.71  &  98.18    &                &      &        & 18.85$\pm$0.05  &         &       2    \\     
2021-01-15   &  230.07  &  99.54    &                &      &   18.70$\pm$0.06     & 18.50$\pm$0.05  &   18.48$\pm$0.04      &       2    \\    
2021-01-16   &  231.10  &  100.57   &    19.92$\pm$0.08	           &  19.21$\pm$0.05    &   19.00$\pm$0.05     & 18.95$\pm$0.03  &   18.72$\pm$0.03      &       2    \\    
2021-01-17   &  232.06  &  101.53   &    20.05$\pm$0.09	           &  19.37$\pm$0.05    &   19.04$\pm$0.04     & 18.95$\pm$0.03  &   18.80$\pm$0.03      &       2    \\    
2021-01-26   &  240.61  &  110.08   &    19.93$\pm$0.15		   &     &   18.95$\pm$0.07     & 18.89$\pm$0.10  &       &       1    \\    
2021-01-26   &  240.63  &  110.10   &                &    &   19.04$\pm$0.12     & 18.89$\pm$0.13  &        &       1    \\    
2021-01-30   &  245.04  &  114.51   &             &  19.22$\pm$0.24    &   18.62$\pm$0.31     & 19.03$\pm$0.15  &   18.93$\pm$0.14      &       2    \\    
2021-02-01   &  246.62  &  116.09   &    20.00$\pm$0.21		   &  19.67$\pm$0.21    &        & 19.18$\pm$0.11  &        &       1    \\    
2021-02-01   &  246.64  &  116.11   &               &  19.67$\pm$0.25    &        & 19.17$\pm$0.11  &         &       1    \\    
2021-02-04   &  250.04  &  119.51   &    19.67$\pm$0.31	           &  19.17$\pm$0.31    &   18.93$\pm$0.22     & 18.91$\pm$0.18  &   18.90$\pm$0.13      &       2    \\    
2021-02-07   &  252.62  &  122.09   &    20.40$\pm$0.31		   &  19.90$\pm$0.29    &   19.28$\pm$0.24     &   &        &       1    \\

\hline    
\multicolumn{9}{l}{\textsuperscript{} $^\dagger$ JD 2,459,000+ ,
$^\ddagger$ Phase has been calculated with respect to $B$$_{max}$ = 2459130.53,
$^\star$ 1: LCO, 2: TNT
} \\
\label{tab:photometric_observational_log_2020udy}                                                        
\end{longtable*} 

The spectroscopic observations of SN~2020udy spanning up to $\sim$ 121 days after the maximum, were obtained with the LCO 2-m Faulkes Telescope North (FTN) in Hawaii, Beijing Faint Object Spectrograph and Camera (BFOSC), mounted on the Xinglong 2.16\,m telescope of NAOC (XLT; \citealt{Fan_2016}), and Dual Imaging Spectrograph (DIS) on the 3.5 m telescope (Astrophysical Research Consortium, ARC) at the Apache Point Observatory. The 1D wavelength and flux-calibrated  LCO spectra were extracted using the \texttt{floydsspec} pipeline\footnote{\url{https://github.com/svalenti/FLOYDSpipeline}} \citep{2014MNRAS.438L.101V}. The wavelength calibration of the BFOSC data and DIS data was done using Fe/Ar and He/Ar arc lamp spectra, respectively. The flux calibration was performed using standard star spectra which had similar airmass as that of the SN. All the spectra were scaled with the photometry to correct for slit loss. Finally, the spectra were corrected for the heliocentric redshift of the host galaxy. The log of spectroscopic observations is given in Table\,\ref{tab:spectroscopic_observations_udy}.

\begin{table*}
\caption{Log of spectroscopic observations of SN 2020udy}
\centering
\smallskip
\begin{tabular}{c c c c c}
\hline \hline
Date     & JD$^\dagger$     & Phase$^\ddagger$             & Spectral Range      & Telescope/Instrument       \\
          &                  &(Days)                             & (\AA)                        &                 \\
\hline
2020-10-01  & 123.87     & -6.6                 & 3300-11000      	  & FTN/FLOYDS  \\
2020-10-02  & 125.20     & -5.3                 & 3500-10000     	  & Xinglong/BFOSC \\
2020-10-04  & 127.19     & -3.3              	& 3500-10000          & Xinglong/BFOSC  \\
2020-10-07  & 129.82     & -0.7                 & 3300-11000          & FTN/FLOYDS  \\
2020-10-09  & 132.21     & 1.68                 & 3500-10000           & Xinglong/BFOSC  \\
2020-10-15  & 137.81     & 7.27                 & 3300-11000          & FTN/FLOYDS \\
2020-10-19  & 142.15     & 11.6                 & 3500-10000          & Xinglong/BFOSC  \\
2020-10-21  & 143.88     & 13.4                 & 3300-11000          & FTN/FLOYDS  \\
2020-10-23  & 144.94     & 14.4             	& 3500-9000           & ARC/DIS\\
2020-10-26  & 148.81     & 18.3                 & 3300-11000          & FTN/FLOYDS             \\
2020-10-26  & 149.17     & 18.6                 & 3500-10000          & Xinglong/BFOSC  \\
2020-10-27  & 150.29     & 19.7                 & 3500-10000          & Xinglong/BFOSC  \\
2020-11-04  & 157.87     & 27.3                 & 3300-11000          & FTN/FLOYDS  \\
2020-11-13  & 166.76     & 36.2                 & 3300-11000          & FTN/FLOYDS  \\
2020-11-14  & 168.20     & 37.7                 & 3500-10000          & Xinglong/BFOSC  \\
2020-11-26  & 179.91     & 49.4                 & 3300-11000          & FTN/FLOYDS  \\
2020-12-03  & 186.89     & 56.4                 & 3300-11000          & FTN/FLOYDS  \\
2020-12-23  & 206.86     & 76.3                 & 3300-11000          & FTN/FLOYDS  \\
2021-01-05  & 219.81     & 89.3                 & 3300-11000          & FTN/FLOYDS   \\
2021-02-06  & 251.72     & 121.2                & 3300-11000          & FTN/FLOYDS   \\

\hline                                   
\end{tabular}
\newline
$^\dagger$JD 2,459,000+
$^\ddagger$ Phase has been calculated with respect to $B$$_{max}$ = 2459130.53
\label{tab:spectroscopic_observations_udy}     
\end{table*}

\section{Distance, extinction and explosion epoch} 
\label{distane_extinction_explosion_epoch}

\subsection{Distance and extinction}
\label{distance and extinction}

There are eight different distance estimates available for the host galaxy of SN~2020udy using Tully-Fisher method \citep{2006Ap.....49..450K,2007A&A...465...71T,2013AJ....146...86T,2014MNRAS.444..527S,2016AJ....152...50T}. Out of these, we have used four recent measurements and scaled them to H$_{0}$ = 73.00 km s$^{-1}$ Mpc$^{-1}$ \citep{Spergel_2007}. The average of these four estimates, 56.20$\pm$12.48 Mpc (consistent with \citealt{2023MNRAS.525.1210M}) is used in this work. 

The SN is located at the outskirts of the host galaxy, so significant reddening from the host galaxy is not expected. Also, in the spectral sequence of SN~2020udy, we do not find the NaID feature associated with the host galaxy. Hence, for reddening correction, we have used only the extinction within the Milky Way, which is $E(B-V)$ = 0.067 mag (A$_{v}$ = 0.210 mag) \citep{2011ApJ...737..103S}. 

\subsection{Explosion epoch}
\label{explosion_epoch}

The first ZTF $r-$band detection at JD 2459116.8 with the last non-detection in $g-$band at 0.87 days ago, together with the comprehensive sampling of SN\,2020udy in ZTF $gr$-bandpasses during a few weeks before its explosion allows a careful inspection of its time of the explosion. Assuming SN\,2020udy exploded as an expanding fireball whose luminosity scales with its surface area, the flux ($F$) thus increases with the square of the time after the explosion (i.e., $F \propto (t-t_{\rm exp})^2$, \citealp{1982ApJ...253..785A, 1999AJ....118.2675R, 2011Natur.480..344N}). Hence, we use Equation~\ref{Eq:ExpEpoch} to fit the extinction-corrected flux in ZTF $g$ and $r$ bands, separately.
\begin{equation}
\label{Eq:ExpEpoch}
    F_{band} = A\times\left(t-t_{exp}\right)^n,
\end{equation}

where A is a scale factor and the power-law index $n$ has been fixed to 2 in the initial fitting, which gives $t_{\rm exp}^{g}=$JD~2459111.9$\pm$0.6 and $t_{\rm exp}^{r}=$JD~2459113.3$\pm$0.4. Such a $\approx$1.4 day discrepancy contradicts the $g-$band last non detection at 2459115.9 JD (see Figure~\ref{fig:SN_2020udy_explosion_epoch}).

Deviation from the $n=2$ expanding fireball during the early-rising phase of type Iax SNe has been reported for multiple cases \citep{2016A&A...589A..89M, 2020ApJ...902...47M}. In the next iteration of fitting, we thus kept $n$ as a free parameter. This resulted in estimates of explosion epochs which were off by 0.5 day between both the bands with  $n$$_{g}$=0.91$\pm$ 0.05 and $n$$_{r}$=1.09$\pm$0.04. 
 
Further, we fit both the light curves iteratively by varying $n$ simultaneously for both the bands in such a way that for the same value of $n$, they converge to the same explosion epoch, which is a free parameter as before.

\begin{figure}
	\begin{center}
		\includegraphics[width=\columnwidth]{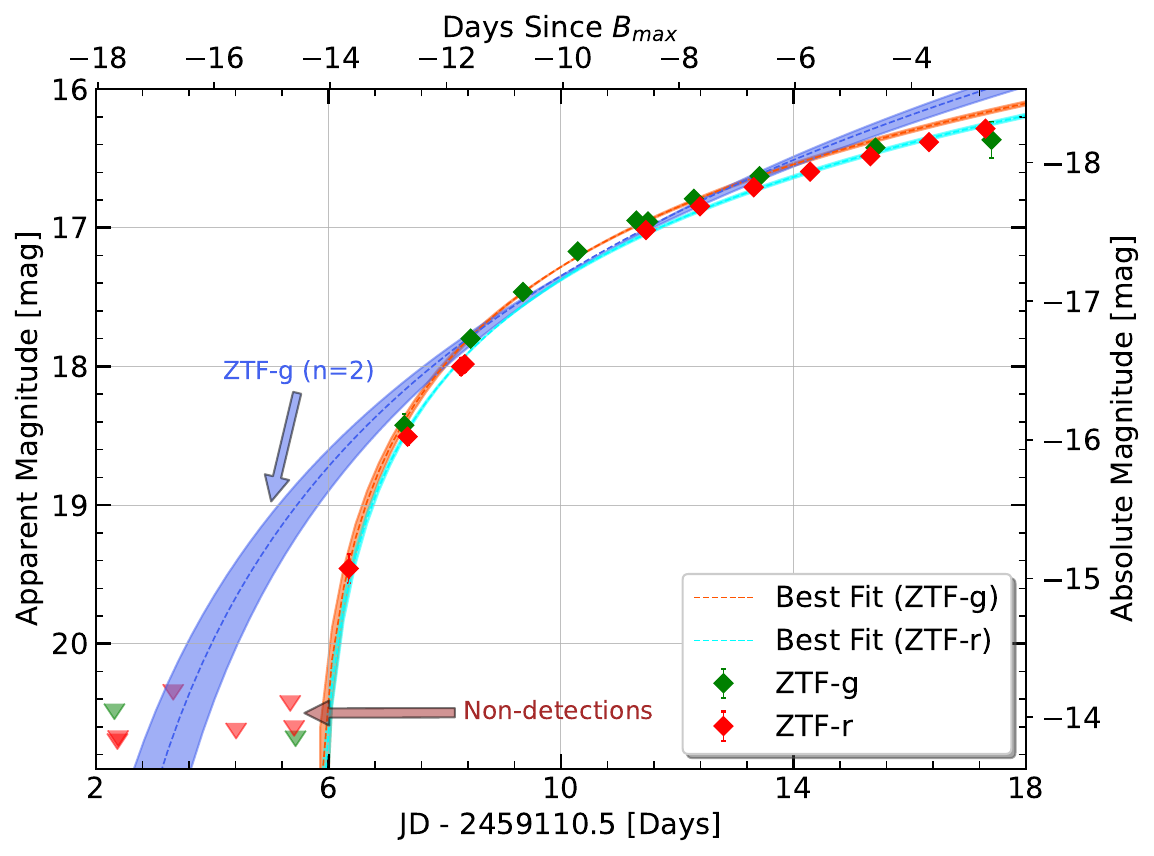}
	\end{center}
	\caption{Estimation of explosion epoch using an analytical expression for the early rise. The 1-$\sigma$ deviations around the best fits are shown with the shaded regions.}
	\label{fig:SN_2020udy_explosion_epoch}
\end{figure}
We found that both the light curves converge to the same explosion epoch ($\rm t_{exp}=2459116.2\pm0.1~JD$) for $n$ = 1.03. \citet{2023MNRAS.525.1210M} quote a similar value for explosion epoch (JD~2459115.7$\pm0.1$), however, they estimated the power index $n$ $\sim$ 1.3 using a modified functional form. In this work, we use JD $2459116.2\pm0.1$ as the explosion date.

\section{Light curve properties}
\label{light_curve_properties}

\subsection{Light curve and color curve}
\label{light_curve_color_curve}

Figure \ref{fig:SN_2020udy_light_curve} presents the {\it BgVri} band photometry of SN~2020udy. The light curves are well sampled around the SN peak brightness in all bandpasses. We use cubic spline fit to estimate the time and magnitude at maximum light in {\it BgVri} bands, the decline in magnitude from the light curve peak to 15 days after (i.e., $\rm \Delta m_{15}$) is also estimated for these bands (Table \ref{tab:lc_param_20udy}).

In Figure 3 we compare the light curves of SN~2020udy with several other well-sampled type Iax SNe, including SNe 2002cx \citep{2003PASP..115..453L}, 2005hk \citep{2008ApJ...680..580S}, 2008ha \citep{2009AJ....138..376F}, 2010ae \citep{2014A&A...561A.146S}, 2011ay \citep{2015A&A...573A...2S}, 2012Z \citep{2015A&A...573A...2S}, 2019gsc \citep{2020ApJ...892L..24S,2020MNRAS.496.1132T}, 2019muj \citep{2021MNRAS.501.1078B}, and 2020rea \citep{2022MNRAS.517.5617S}. The light curves are 
normalized to their peak magnitudes in the respective bands and plotted in the rest frame of individual SN. In {\it B} band, SN~2020udy declines slower than SNe 2002cx and 2011ay and its light curve shape looks remarkably similar to that of SNe 2012Z and 2020rea. In {\it g} band, SN~2020udy declines slower than all the comparison SNe. The {\it V} band light curve evolution of SN~2020udy is faster than SNe\,2002cx, 2005hk, 2011ay, 2012Z, 2020rea and slower compared to the other SNe presented in our sample. With a slower decline in {\it r} and {\it i} bands, SN~2020udy appears to be very similar to SNe 2012Z and 2020rea (Figure \ref{fig:SN_2020udy_comp_light_curve}).

The {\it B-V}, {\it V-I}, {\it V-R}, and {\it R-I}} color evolution of SN~2020udy and its comparison with other type Iax SNe have been depicted in Figure \ref{fig:SN_2020udy_color_curve}. All the colors are corrected for total reddening and brought to the rest frame of each SN. The color evolution of SN~2020udy seems to follow the same pattern as other type Iax SNe used for comparison.

\begin{table*}
\caption{ Light curve parameters of SN 2020udy  }
\centering
\smallskip
\begin{tabular}{l  c c c c c}
\hline \hline
SN 2020udy                          & B band           & g band           & V band            & r band              & i band  \\
\hline
JD of maximum light (2459000+)      & 130.53$\pm$1.0   & 130.54$\pm$1.0   & 130.54$\pm$1.0    & 133.74$\pm$1.0      &139.93$\pm$1.0     \\
Magnitude at maximum (mag)          & 16.72$\pm$0.02   & 16.46$\pm$0.01   & 16.40$\pm$0.01    & 16.18$\pm$0.03      & 16.25$\pm$0.01    \\
Absolute magnitude at maximum (mag) &-17.41$\pm$0.34   & -17.56$\pm$0.34  & -17.76$\pm$0.34   & -17.83$\pm$0.34     & -17.82$\pm$0.33    \\
$\Delta$m$_{15}$( mag)              & 1.36$\pm$0.04    & 1.18$\pm$0.04    & 0.45$\pm$0.04     & 0.49$\pm$0.03       & 0.66$\pm$0.04   \\ 
\hline

\end{tabular}
\newline
\label{tab:lc_param_20udy}      
\end{table*}

\begin{figure}
	\begin{center}
		\includegraphics[width=\columnwidth]{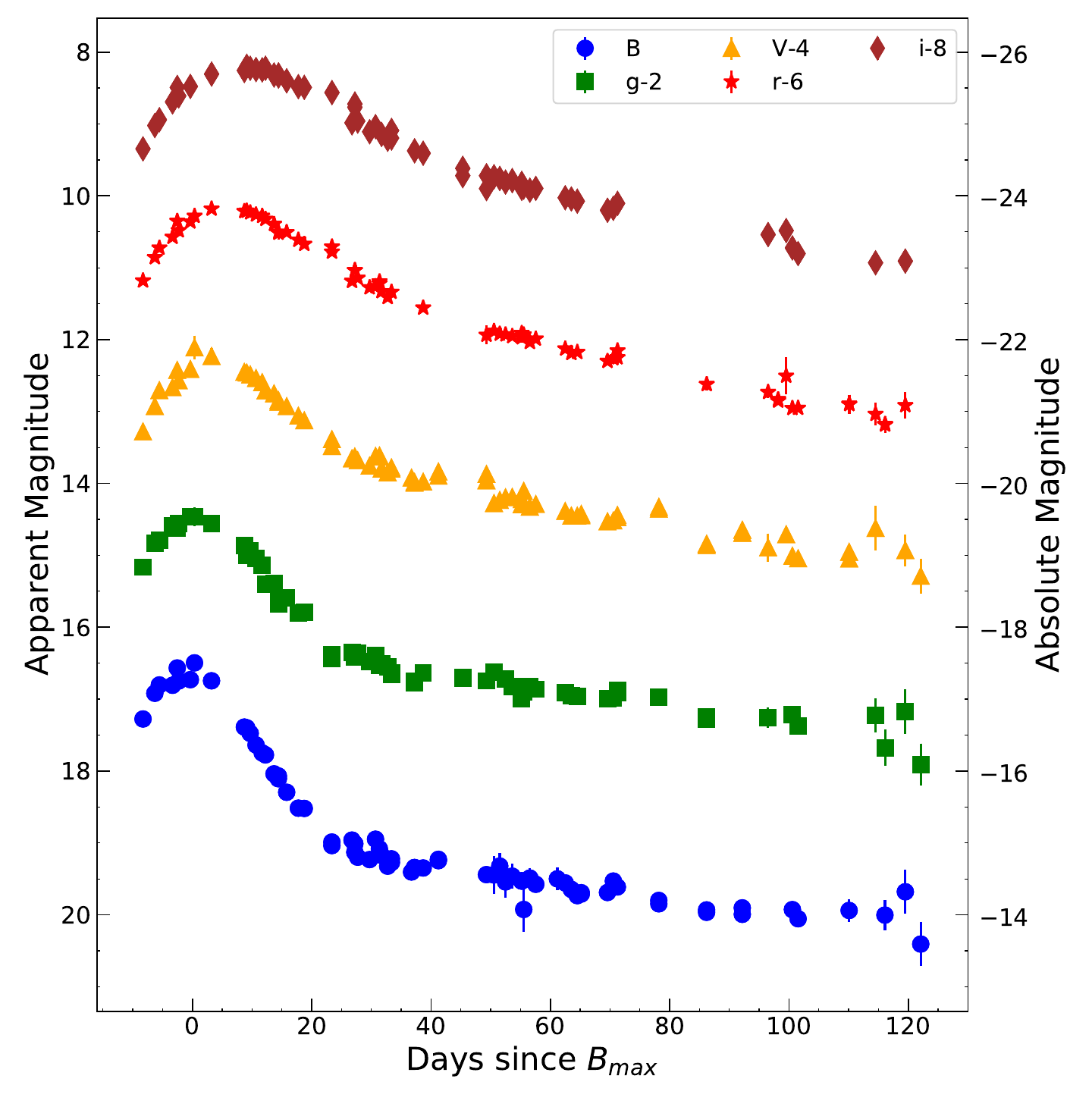}
	\end{center}
	\caption{Light curve evolution of SN~2020udy in {\it BgVri} bands.}
	\label{fig:SN_2020udy_light_curve}
\end{figure}

\begin{figure}
	\begin{center}
		\includegraphics[width=\columnwidth]{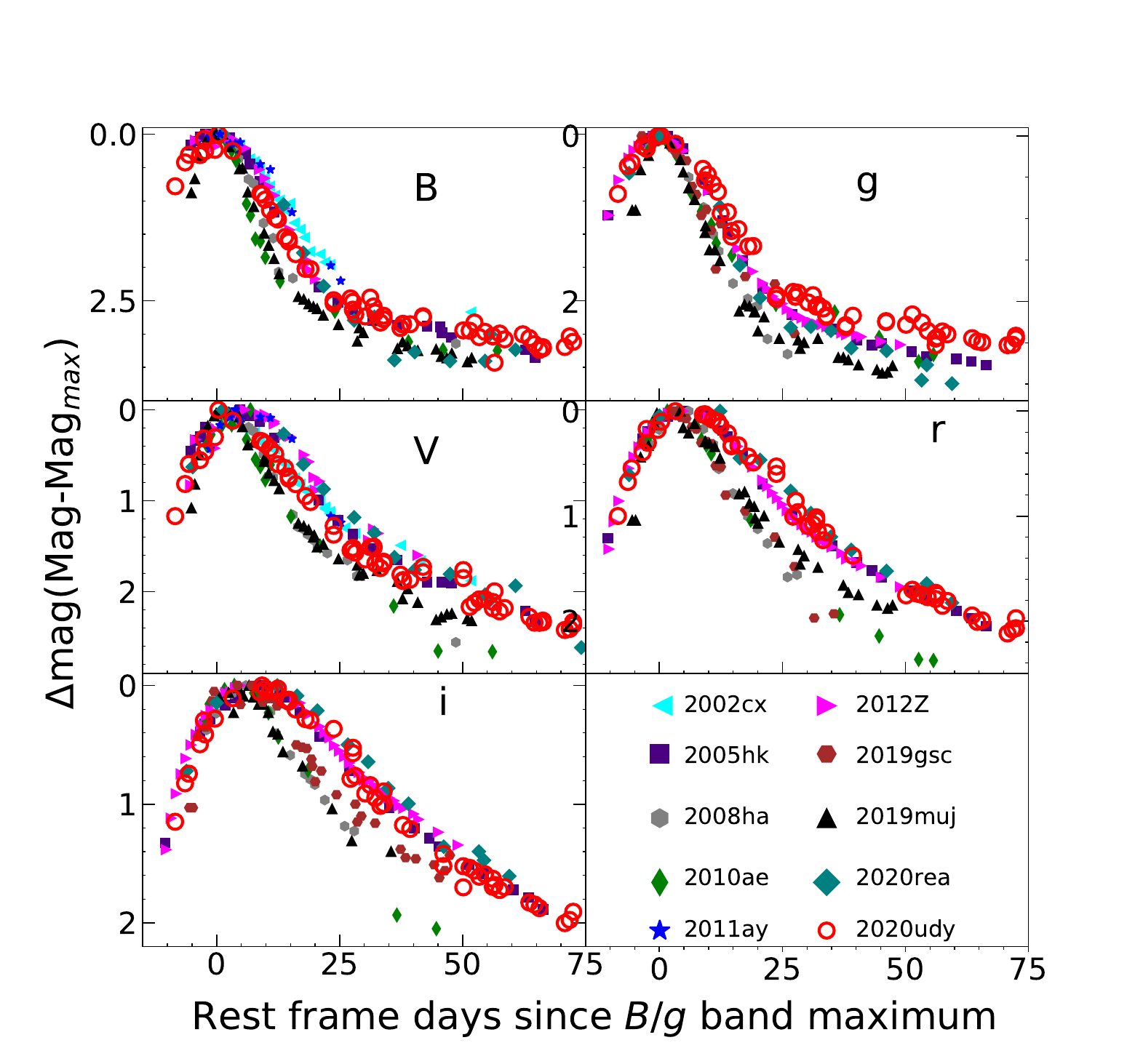}
	\end{center}
	\caption{Light curve comparison of SN~2020udy with other type Iax SNe. The date of maximum in {\it B} band is used as reference for comparison plots in {\it B} and {\it V} bands while comparison plots in {\it gri} bands are with respect to {\it g} band maximum.}
	\label{fig:SN_2020udy_comp_light_curve}
\end{figure}

\begin{figure}
	\begin{center}
		\includegraphics[width=\columnwidth]{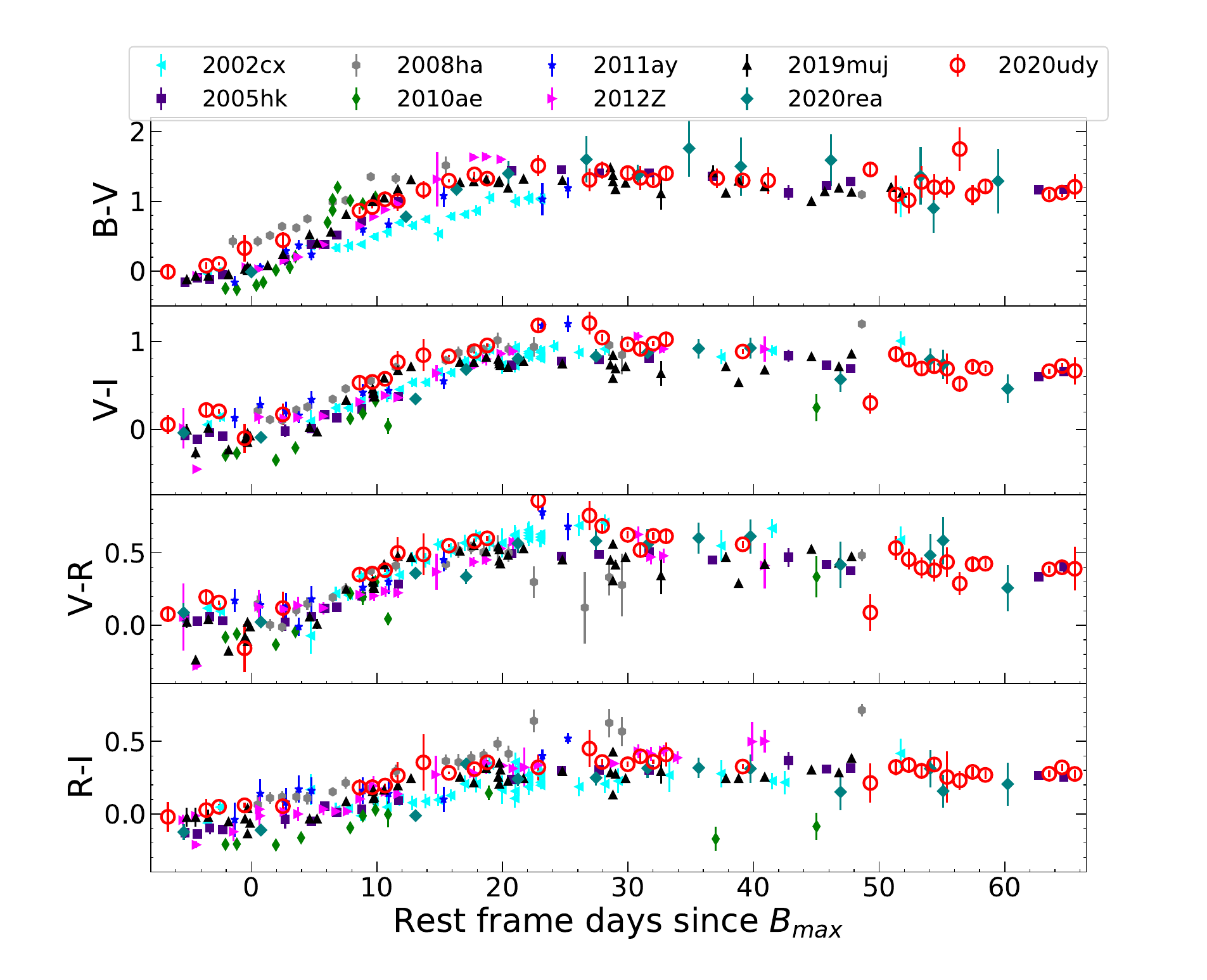}
	\end{center}
	\caption{Color evolution of SN~2020udy with those of other type Iax SNe.}
	\label{fig:SN_2020udy_color_curve}
\end{figure}

\subsection{Analysis of bolometric light curve}
\label{bolometric_light_curve}

\begin{figure}
	\begin{center}
		\includegraphics[width=\columnwidth]{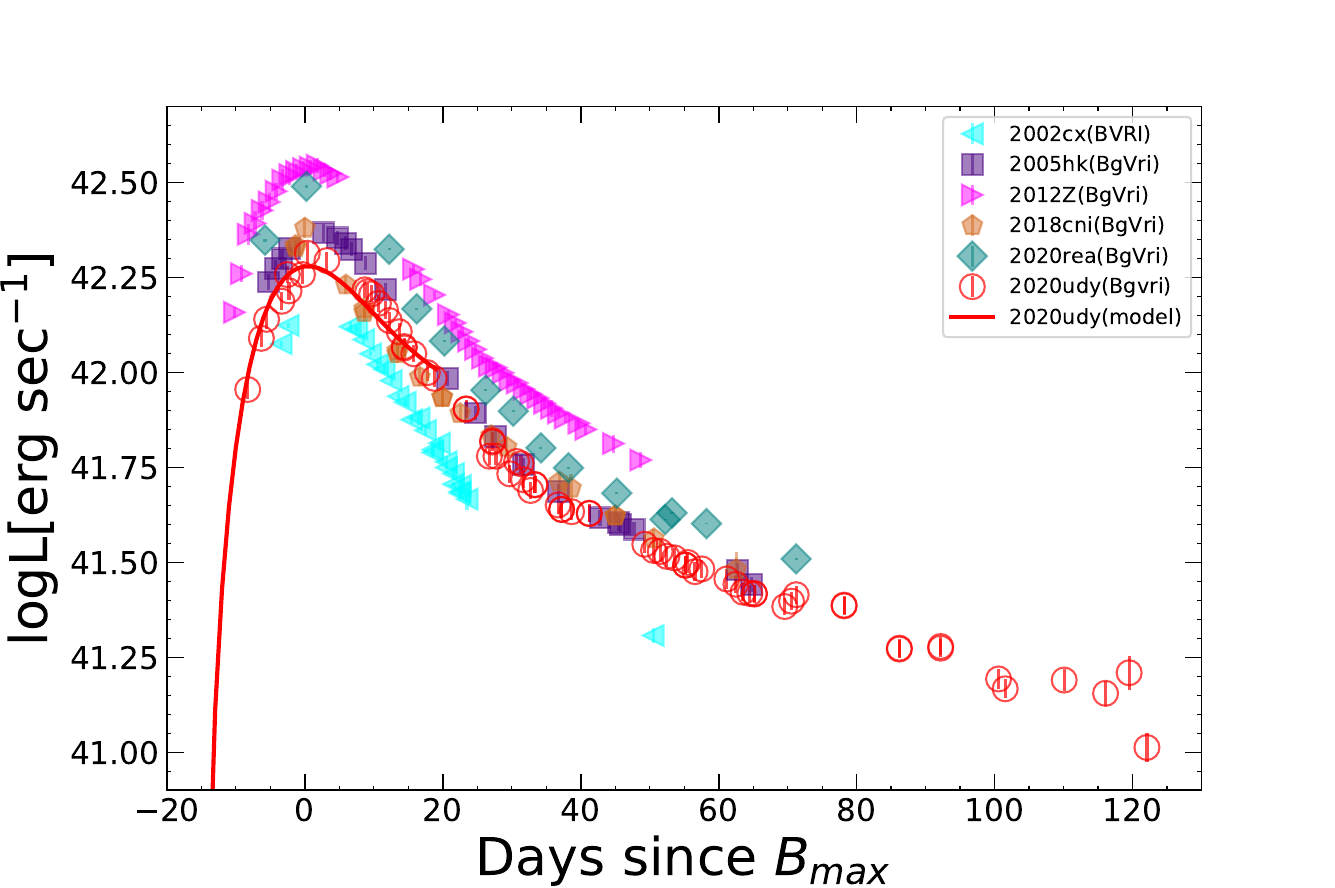}
	\end{center}
	\caption{This figure presents the pseudo bolometric light curve of SN~2020udy with those of other type Iax SNe. The analytical model used to estimate the explosion parameters is also shown by a solid line.}
	\label{fig:SN_2020udy_bolometric}
\end{figure}

The pseudo-bolometric {\it BgVri} light curve of SN~2020udy is constructed using \texttt{SuperBol} \citep{2018RNAAS...2..230N}, adopting distance and extinction discussed in Section \ref{distance and extinction}. In \texttt{SuperBol}, the dereddened magnitudes and associated errors of the SN are converted to fluxes and flux errors. These fluxes are used to construct the spectral energy distribution (SED) at all epochs, which are integrated using trapezoidal rule in the limit of the input passbands to obtain the pseudo-bolometric luminosity. The flux errors and bandwidths are used to calculate the corresponding pseudo-bolometric luminosity error. The integration spans the wavelength range between 3960 and 8120 \AA. To estimate the missing flux in UV and NIR region, \texttt{SuperBol} fits blackbody functions to SED at each epoch and applies this correction to the pseudo-bolometric luminosities to construct the total bolometric luminosity. Figure \ref{fig:SN_2020udy_bolometric} shows the pseudo-bolometric light curve of SN~2020udy along with a few other bright type Iax SNe. The pseudo-bolometric light curves of all the SNe presented in Figure \ref{fig:SN_2020udy_bolometric} are constructed similarly. We have added another bright type Iax SN 2018cni \citep{2023ApJ...953...93S} for comparison in Figure \ref{fig:SN_2020udy_bolometric}. Around maximum, SN~2020udy looks slightly fainter than SNe 2005hk and 2018cni but after $\sim$ 10 days their pseudo-bolometric luminosities are comparable.  This shows that SN~2020udy is a bright type Iax SN with peak ({\it BgVri}) luminosity 2.06$\pm$0.14 $\times$10$^{42}$ erg sec$^{-1}$. 

To estimate the explosion parameters, such as the mass of $^{56}$Ni and ejecta mass (M$_{ej}$), we employed an analytical model proposed by \cite{1982ApJ...253..785A} and \cite{2008MNRAS.383.1485V}. This model assumes spherically symmetric and optically thick ejecta, a small initial radius, constant optical opacity, and the presence of $^{56}$Ni in the ejected matter. We fitted the pseudo bolometric light curve with the analytical model using scipy.optimize.curve\_fit algorithm, setting $\kappa_{opt}$ to 0.1\,cm$^{2}$\,g$^{-1}$ and the photospheric velocity to 8000\,km\,s$^{-1}$ at maximum. From the fitting, we obtained the following values of the parameters: $^{56}$Ni mass = 0.08$\pm$0.01\,M$_{\odot}$ and ejecta mass (M$_{ej}$) = 1.39$\pm$0.09\,M$_{\odot}$, with the errors obtained from the covariance matrix.

The bolometric rise time obtained through this fit is $\sim$ 15 days. This is consistent with the rise time estimated independently by fitting the early light curves of SN~2020udy (see, Section \ref{explosion_epoch}).

To obtain the total bolometric luminosity from the pseudo-bolometric luminosity, the contribution from the missing bands needs to be added. For type Iax SNe, a definite contribution of UV and NIR fluxes to the total bolometric luminosity is not known, as for only a handful of objects, both UV and NIR coverage are available in literature \citep{2007PASP..119..360P,2015ApJ...806..191Y,2016MNRAS.459.1018T,2020ApJ...892L..24S,2022ApJ...925..217D,2022MNRAS.511.2708S}. Based on the available estimates, if the contribution from UV and NIR bands to the total luminosity at maximum is taken as $\sim$ 35\%, the mass of $^{56}$Ni increases to 0.11 M$_{\odot}$.

\begin{figure}
	\begin{center}
		\includegraphics[width=\columnwidth]{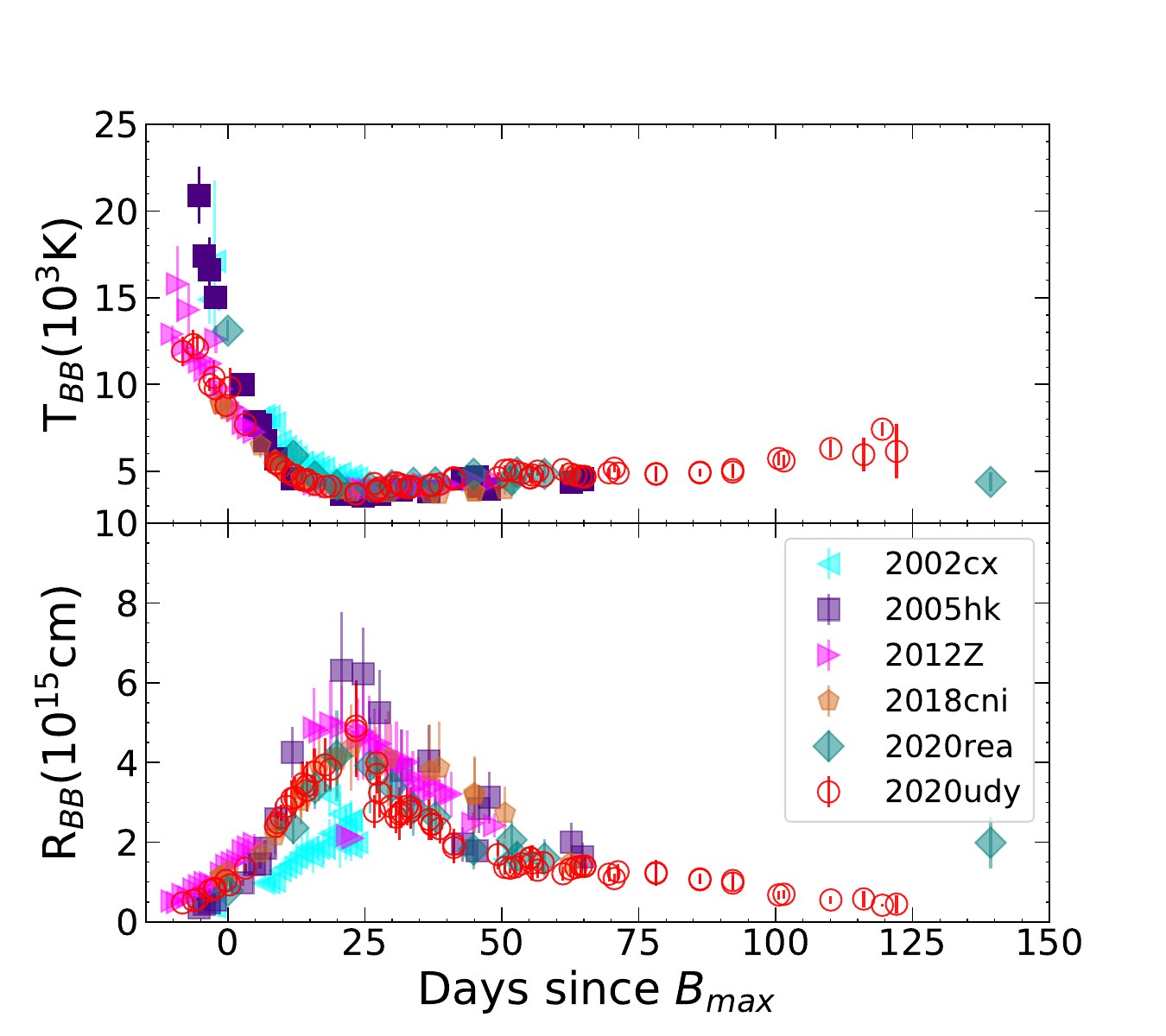}
	\end{center}
	\caption{Evolution of blackbody temperature ($T_{BB}$, upper panel) and radius ($R_{BB}$, lower panel) of SN~2020udy compared to that of several other bright type Iax SNe. }
	\label{fig:SN_2020udy_bolometric_radtemp}
\end{figure}

Figure \ref{fig:SN_2020udy_bolometric_radtemp} compares the evolution of blackbody temperature and radius of SN 2020udy with that estimated for several other well-studied type Iax SNe. The blackbody temperature evolves in a similar fashion for all the type Iax SNe presented in Figure \ref{fig:SN_2020udy_bolometric_radtemp}, whereas the blackbody radius is proportional to the luminosity of the SN.

\begin{figure}
	\begin{center}
		\includegraphics[width=\columnwidth]{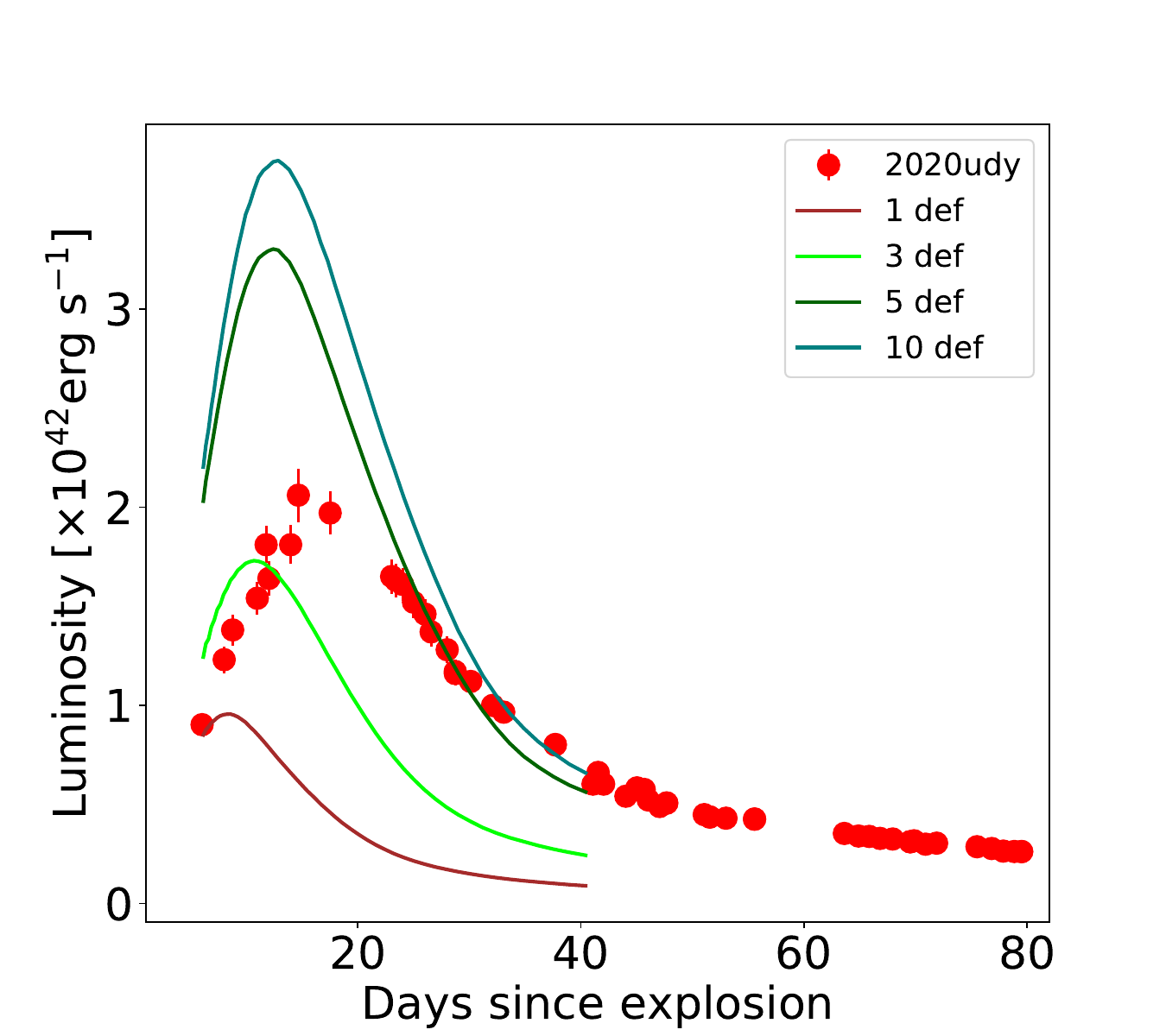}
	\end{center}
	\caption{Comparison of the pseudo bolometric light curve of SN~2020udy with optical bolometric luminosity evolution of different} deflagration models presented in \citet{2014MNRAS.438.1762F}.
	\label{fig:SN_2020udy_deflagration}
\end{figure}

In Figure \ref{fig:SN_2020udy_deflagration}, we compare the pseudo-bolometric luminosity of SN~2020udy with deflagration models presented in \cite{2014MNRAS.438.1762F}. The model light curves N1-def, N3-def, N5-def, and N10-def are adopted from the HESMA database. The numerical value indicates number of the ignition spots of the model, which approximately scales with the strength of the deflagration. During the photospheric phase, the pseudo-bolometric luminosity of SN~2020udy lies between N3-def and N5-def models. Around 30 days post-explosion, SN~2020udy shows a slower decrease in the bolometric luminosity as compared to all the models presented in Figure \ref{fig:SN_2020udy_deflagration}, which may indicate a higher ejecta mass of SN\,2020udy compared to the model predictions.

\section{Spectral evolution}
\label{spectral evolution}

\subsection{Spectral features of SN~2020udy and their comparison with other type Iax SNe }
\label{spectral_features_and_comparison}

\begin{figure}
	\begin{center}
		\includegraphics[width=\columnwidth]{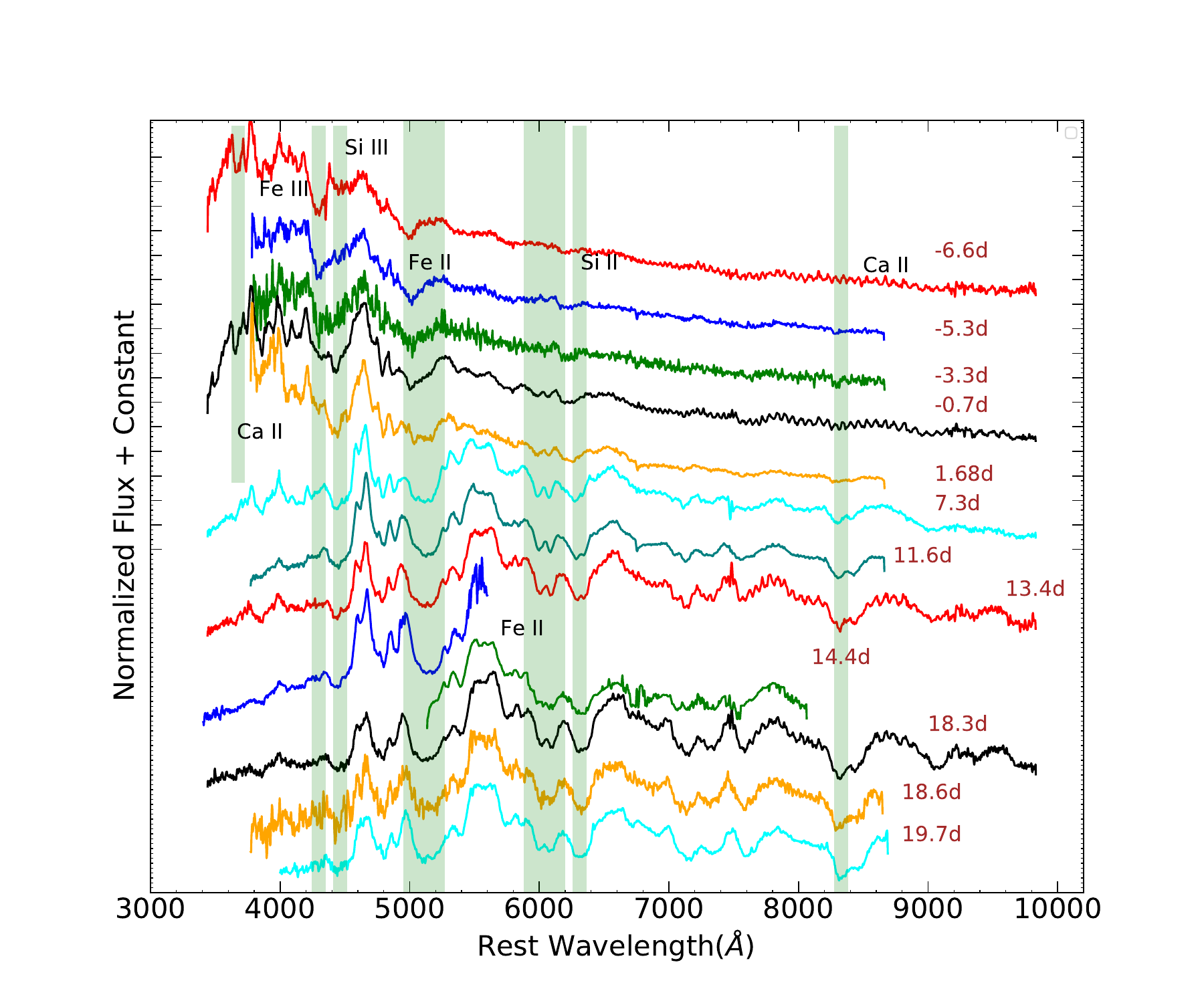}
	\end{center}
	\caption{Spectral evolution of SN~2020udy from -6 days to $\sim$20 days since the maximum.}
	\label{fig:SN_2020udy_spectra_plot1}
\end{figure}

\begin{figure}
	\begin{center}
		\includegraphics[width=\columnwidth]{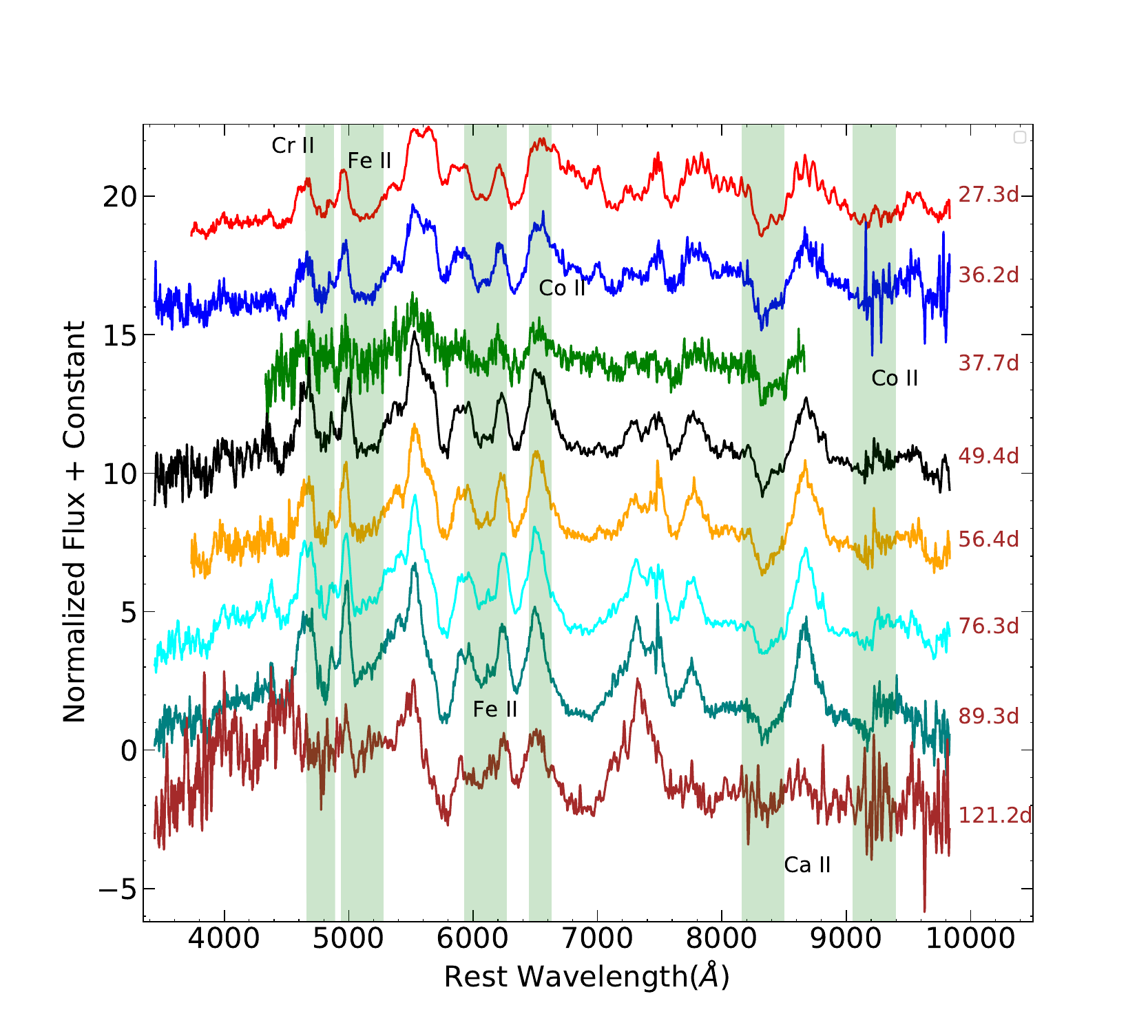}
	\end{center}
	\caption{Spectral evolution of SN~2020udy spanning between 27 to 121 days since the maximum.}
	\label{fig:SN_2020udy_spectra_plot2}
\end{figure}

\begin{figure}
	\begin{center}
		\includegraphics[width=\columnwidth]{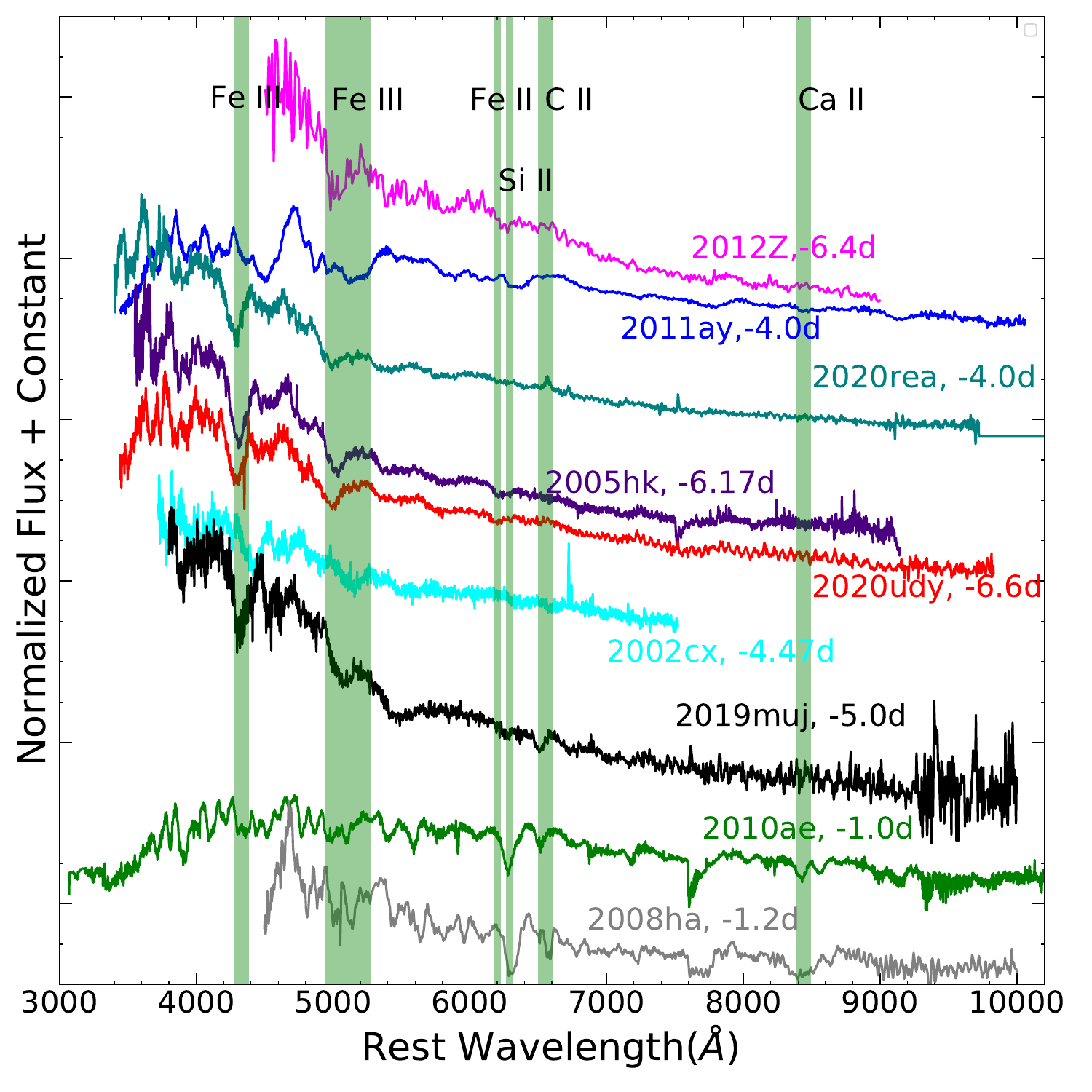}
	\end{center}
	\caption{Pre-peak spectral features of SN~2020udy compared with those of other type Iax SNe at similar phases. The spectra are plotted in the decreasing order of $B-$band peak brightness of the SNe.}
	\label{fig:pre_peak_comp_spectra_plot1}
\end{figure}

Figures \ref{fig:SN_2020udy_spectra_plot1} and \ref{fig:SN_2020udy_spectra_plot2} present the spectroscopic evolution of SN~2020udy spanning from $-$6.6 to $+$121.2 days relative to its {\it B}-band maximum light. During the pre-maximum phase, the spectra of SN~2020udy show a blue continuum with features due to Ca {\sc II} H\&K, Fe {\sc III}, Si {\sc III}, Fe {\sc II}, and Si {\sc II}. Prominent P-Cygni profiles with a broad absorption component can also be identified. Similar to other bright type Iax SNe, SN~2020udy also exhibits a rather shallow Si\,{\sc II} absorption line at 6355 \AA. We do not detect the C {\sc II} feature at 6580 \AA. Figure \ref{fig:pre_peak_comp_spectra_plot1} presents a comparison of the earliest spectrum of SN~2020udy at  $-$6.6 days with several other type Iax SNe observed at similar phases. Fe {\sc III} features near 4400 and 5000 \AA~ are present in all the SNe. Additionally, we remark that the strength of both Si {\sc II} and C {\sc II} features diminishes with the increasing luminosity of type Iax SNe. For example, fainter events like SNe 2008ha and 2010ae display significantly stronger Si {\sc II} and obvious C {\sc II} absorption lines compared to bright events such as SNe\,2012Z, 2020rea, 2020udy etc. Similarly, in faint type Iax SNe, Ca {\sc II} NIR feature also emerges earlier than in the bright type Iax SNe. 

\begin{figure}
	\begin{center}
		\includegraphics[width=\columnwidth]{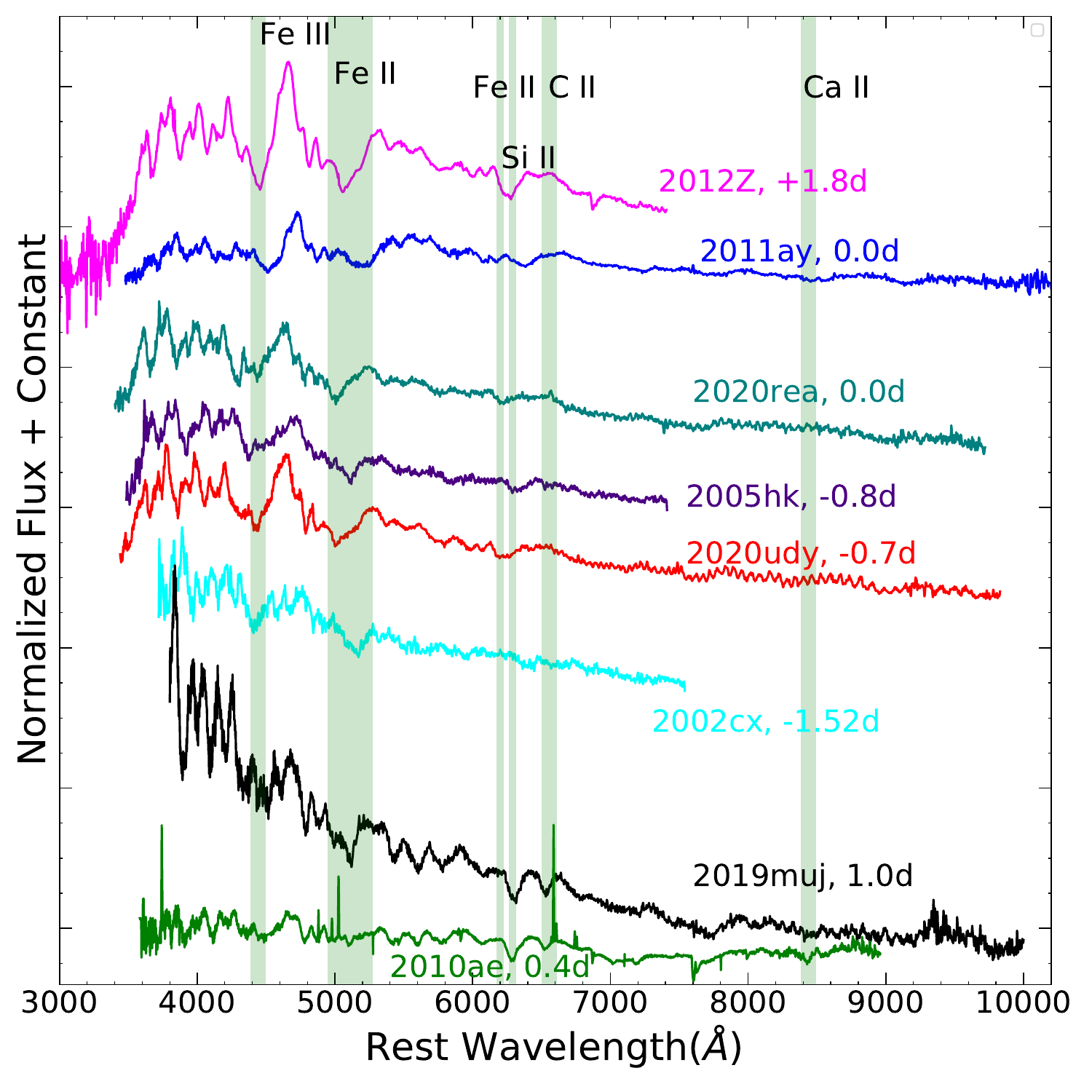}
	\end{center}
	\caption{Comparison between the spectrum of SN~2020udy around the $B-$band peak brightness with other type Iax SNe observed at similar phases.}
	\label{fig:peak_comp_spectra_plot2}
\end{figure}

Comparison of spectral features near maximum (Figure \ref{fig:peak_comp_spectra_plot2}) indicates similarity between SN~2020udy and other bright type Iax SNe. Near maximum, the strength of Si {\sc II}  6355 \AA~ line increases. Features due to  Fe {\sc III} near 4400 \AA, Fe {\sc II} near 5000 \AA, Si {\sc II}, and Ca {\sc II} NIR triplet are prominently seen in all the SNe. In the early post-maximum phase ($\sim$ 7 days), the blue part of the spectrum gets suppressed because of cooling of the ejecta and line blanketing (Figure \ref{fig:SN_2020udy_spectra_plot1}). With time, the Si {\sc II} line is replaced by the progressively emerging Fe/Co lines (Figure \ref{fig:SN_2020udy_spectra_plot2}). The features between 5500 to 7000 \AA~ are mostly dominated by Fe II lines. 

\begin{figure}
	\begin{center}
		\includegraphics[width=\columnwidth]{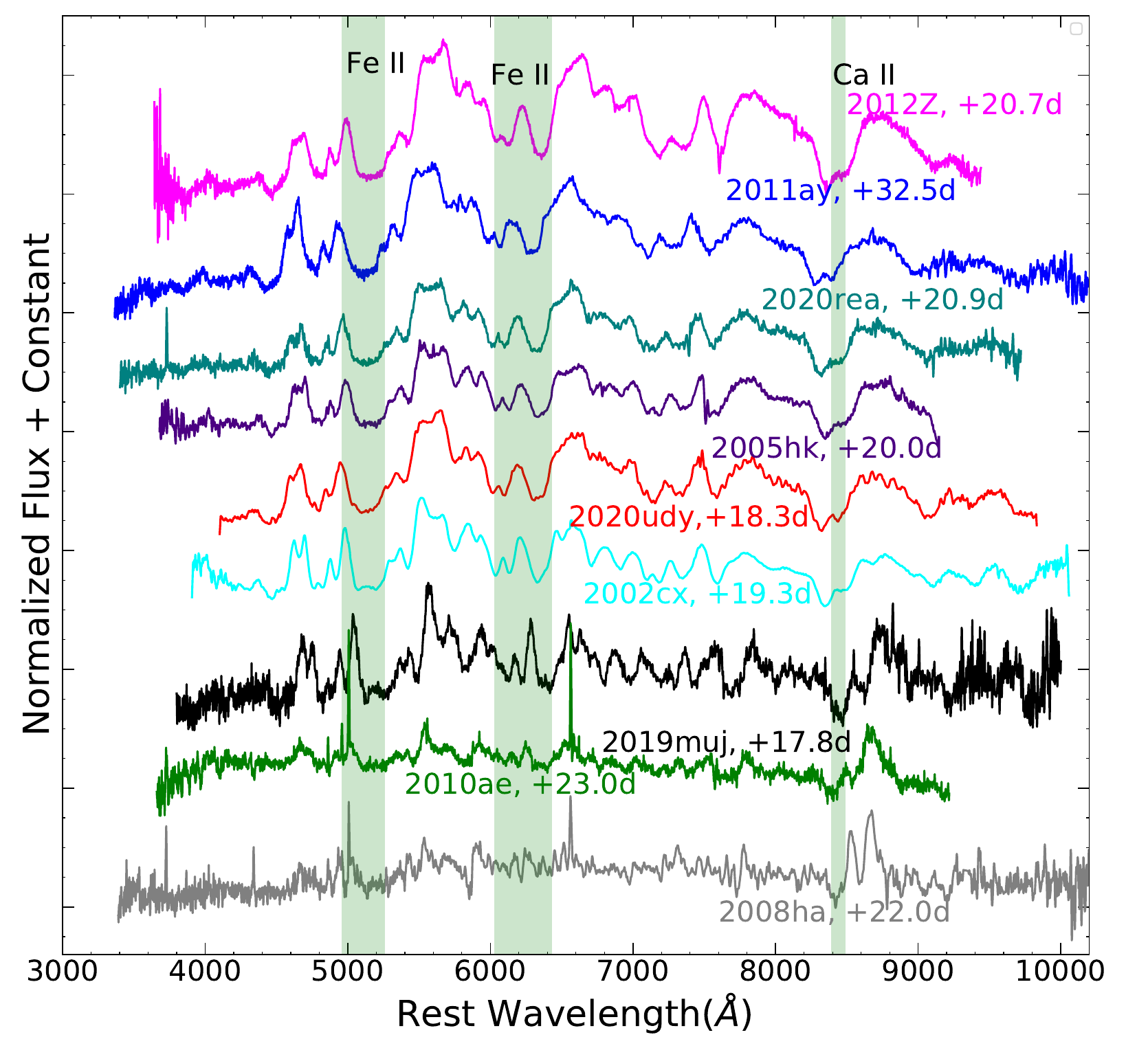}
	\end{center}
	\caption{Comparison of post-peak spectral features of SN~2020udy with other type Iax SNe.}
	\label{fig:post_peak_comp_spectra_plot}
\end{figure}

Post maximum (after $\sim$ 20 days since maximum) spectral evolution of SN 2002udy is shown in Figure \ref{fig:SN_2020udy_spectra_plot2}. Cr\,{\sc II} feature near 4800 \AA~ and Co\,{\sc II} feature near 6500 \AA~ are clearly visible. Ca {\sc II} NIR triplet becomes progressively stronger. Figure \ref{fig:post_peak_comp_spectra_plot} compares the spectrum of SN~2020udy obtained at day $+$18.3 with the spectra of other type Iax SNe at similar phases. The spectra of all SNe in the sample are dominated by Iron Group Elements (IGEs). SN~2020udy exhibits remarkable similarities to bright type Iax SNe such as SNe 2005hk, 2011ay, and 2020rea. An absorption feature at $\sim$9000\,\AA, due to Co {\sc II} is also seen in SN~2020udy. As the inner ejecta of the SN becomes optically thin during the late phase, emission lines increase in strength.

\begin{figure}
	\begin{center}
		\includegraphics[width=\columnwidth]{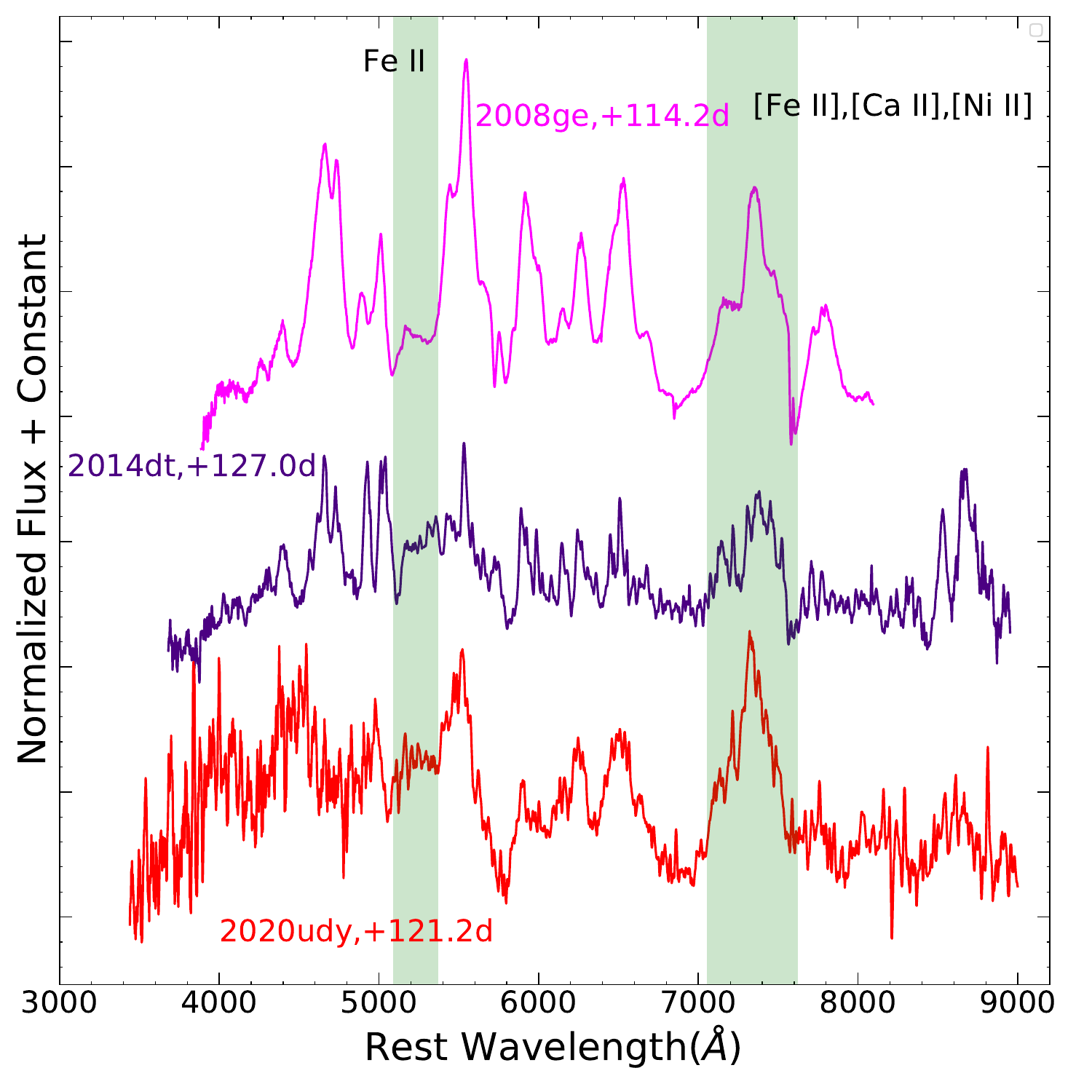}
	\end{center}
	\caption{Comparison of nebular phase spectral features of SN~2020udy with other type Iax SNe.}
	\label{fig:nebular_phase_comp_spectra_plot}
\end{figure}

In the late phase, spectral features become narrow (Figure \ref{fig:SN_2020udy_spectra_plot2}). The region around 7300 \AA~ is composed of forbidden lines of Fe/Ni and Ca \citep{2016MNRAS.461..433F}. The presence of both forbidden and permitted lines in the late phase spectra of SN~2020udy indicates that the spectrum is not fully nebular. Type Iax SNe posses a long-lived photosphere with the presence of permitted lines even at late times \citep{2006AJ....131..527J,2008ApJ...680..580S,2010ApJ...708L..61F,2016MNRAS.461..433F}. Figure \ref{fig:nebular_phase_comp_spectra_plot} shows nebular phase spectral comparison between SNe 2020udy, 2008ge \citep{2010AJ....140.1321F}, and 2014dt \citep{2018MNRAS.474.2551S} at comparable epoch. The nebular phase spectrum of SN~2020udy and SN~2008ge \citep{2010AJ....140.1321F} shows broad emission feature, while SN\,2014dt has narrow spectral features. \cite{2023MNRAS.525.1210M} have also reported broad emission features in the spectrum of SN 2020udy at $\sim$ 119 and 137 days and suggested that it might be coming from the SN ejecta. In the case of SN 2012Z, similar broad emission features were reported at a very late phase ($\sim$ 190 days, \citealt{2015A&A...573A...2S}). \citet{2016MNRAS.461..433F} suggested that relatively bright type Iax SNe with higher ejecta velocities exhibit broad forbidden lines; SN~2020udy is consistent with these findings.       

\subsection{Evolution of photospheric velocity}
\label{velocity_evolution}

\begin{figure}
	\begin{center}
		\includegraphics[width=\columnwidth]{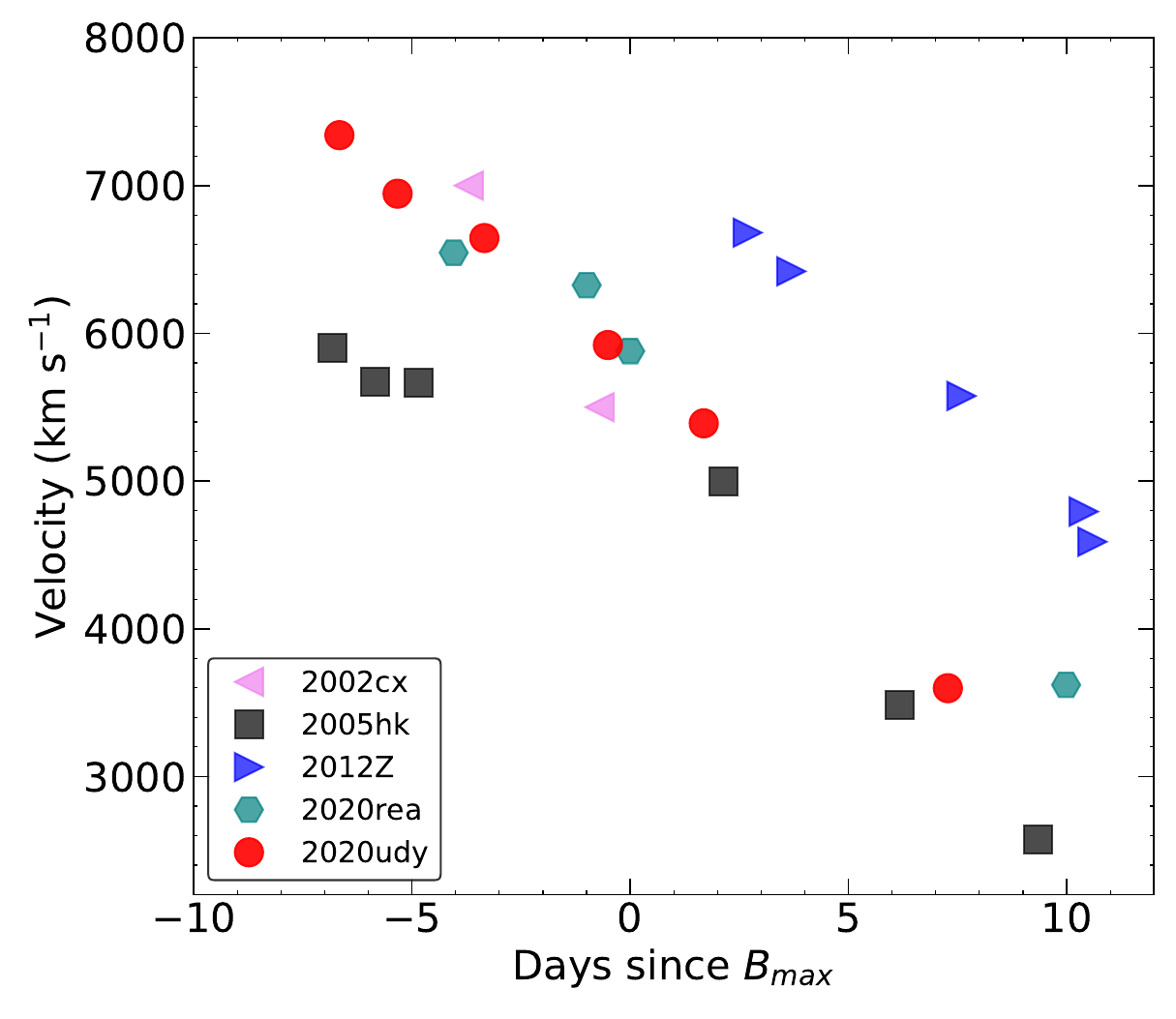}
	\end{center}
	\caption{Evolution of the velocity of Si {\sc II} line in SN~2020udy with those of other types Iax SNe is shown in this figure.}
	\label{fig:SN_2020udy_velocity}
\end{figure}

Figure \ref{fig:SN_2020udy_velocity} displays the photosphere expansion velocity of SN~2020udy traced by the evolution of the Si {\sc II} line at 6355 \AA~ and that measured for several other type Iax SNe. We estimate the expansion velocity by fitting a Gaussian profile to the absorption minimum of the Si {\sc II} line. The expansion velocity of SN~2020udy around maximum light is  $\sim$ 6000 km s$^{-1}$. During $-6.6$ and $+7.27$ days, the photospheric velocity of SN~2020udy is similar to SN 2020rea, higher than SN 2005hk, and lower than SN 2012Z. The photospheric velocity estimated by the absorption minimum of the Si {\sc}{II} line at  6355 \AA~ becomes unreliable after $\sim$two weeks relative to the $B-$band peak brightness due to the increasing blending with the emerging iron lines. Type Iax SNe having higher luminosity are known to have high photospheric velocity \citep{2010ApJ...720..704M,2013ApJ...767...57F} except for a few outliers. We found that SN~2020udy is consistent with such a luminosity-velocity correlation.

\subsection{Spectral modelling}
\label{spectral_modelling}

\subsubsection{Spectral modeling with \texttt{SYNAPPS}}
\label{spectral_modelling_synapps}

\begin{figure}
	\begin{center}
		\includegraphics[width=\columnwidth]{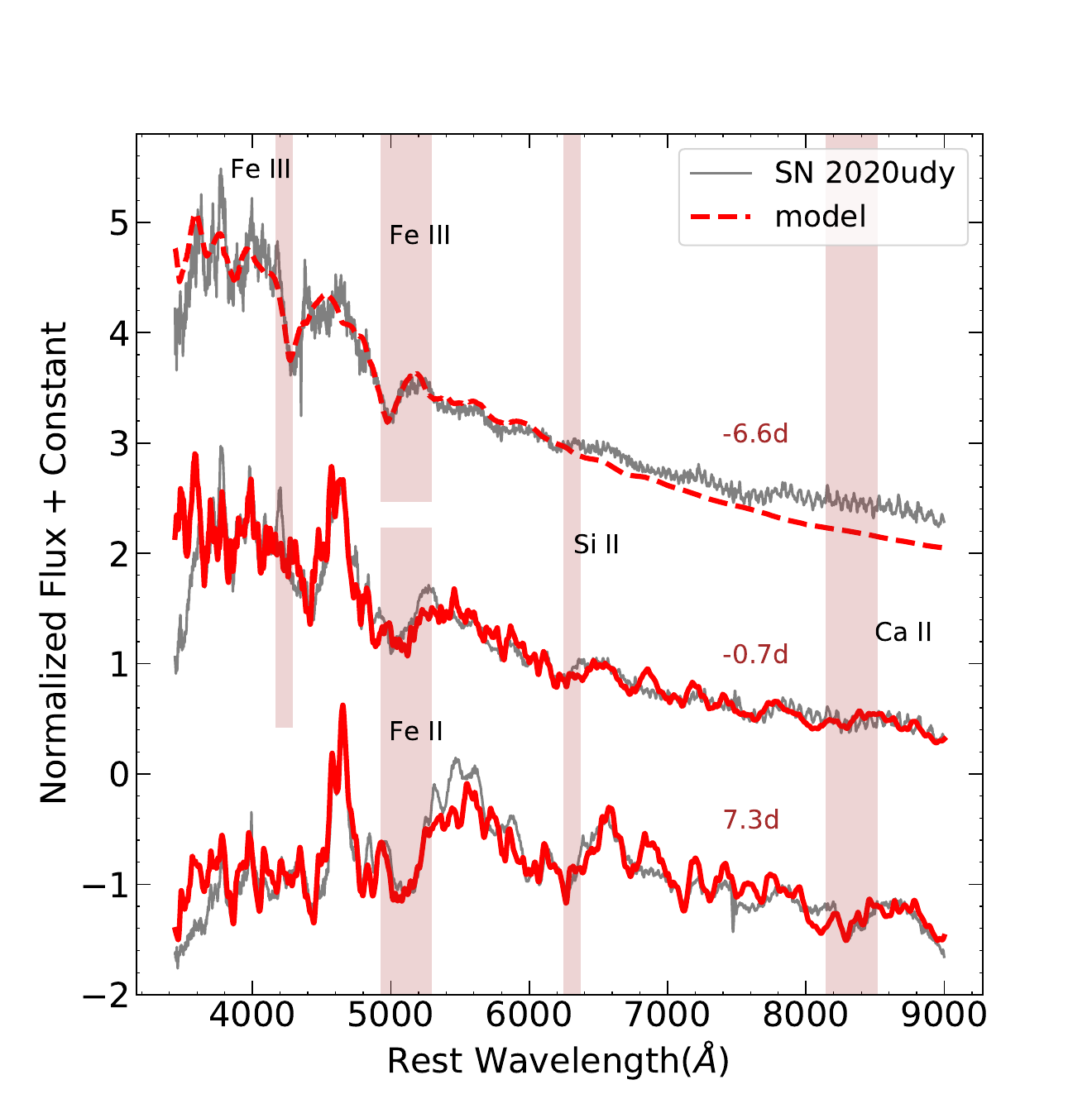}
	\end{center}
	\caption{The $-$6.6, $-$0.7, and 7.3 days spectra of SN~2020udy are shown (in grey) along with the respective SYNAPPS models (in red).}
	\label{fig:SN_2020udy_synapps}
\end{figure}

Photospheric spectra of SN~2020udy at three epochs: $-$6.6, $-$0.7, and 7.3 days relative to the $B-$band peak brightness are modeled using the spectrum synthesis code  \texttt{SYNAPPS} \citep{2011PASP..123..237T} and are shown in Figure \ref{fig:SN_2020udy_synapps}. The photospheric velocity in the best-fit model falls from 11250 km s$^{-1}$ at $-$6.6 days to 6570 km s$^{-1}$ at 7.3 days. The outer ejecta velocity used in the modeling is 30000 km s$^{-1}$. The photospheric temperature evolves from  12200 K at $-$6.6 days to  8500 K at 7.3 days. The chemical species used in the modeling are Fe, Si, Ca, Mg, and S. Specifically, in the pre-maximum spectral fitting, we used S {\sc II}, Si {\sc II}, Si {\sc III}, Mg {\sc II}, Ca {\sc II}, Fe {\sc II} and Fe {\sc III} ions. The prominent Fe {\sc II} and Fe {\sc III} features in the pre-maximum spectrum are reproduced very well in the models. In the model spectra at maximum and post maximum, most of the spectral features such as Fe, Si, Ca, etc. match well with our observed spectra. 

\subsubsection{Spectral modeling with TARDIS}
\label{spectral_modelling_tardis}

We perform spectral modeling for SN~2020udy using the one-dimensional radiative transfer code
TARDIS, following the principles of abundance tomography. The setting of TARDIS and the fitting strategy is almost the same as in our previous study, where SNe 2018cni and 2020kyg were the subjects of analysis \citep{2023ApJ...953...93S}. The method was previously applied for abundance tomography of normal SNe Ia using Artificial Intelligence-Assisted Inversion \citep[AIAI,][]{2020ApJS..250...12C,2022arXiv221015892C} and fit-by-eye method \citep{2021MNRAS.506..415B} as well, providing similar abundance profiles and the same goodness of fit.

The present spectral synthesis covers a longer range of time, $\sim90$ days after the maximum. \cite{2023ApJ...951...67C} showed that radiative transfer codes assuming a blackbody-emitting photosphere can reproduce most of the spectral features and their evolution over years after the explosion because type Iax SNe never show a fully nebular spectrum. At the same time, spectral synthesis is less sensitive to the exact mass fractions of chemical elements at $t_\mathrm{exp} > 30$ days, compared to the earlier epochs. The information is retrievable from late-time spectral fitting limits to luminosity ($L$), photospheric velocity ($v_\mathrm{phot}$), and identification of the presence of chemical elements.

\begin{figure}
	\begin{center}
		\includegraphics[width=\columnwidth]{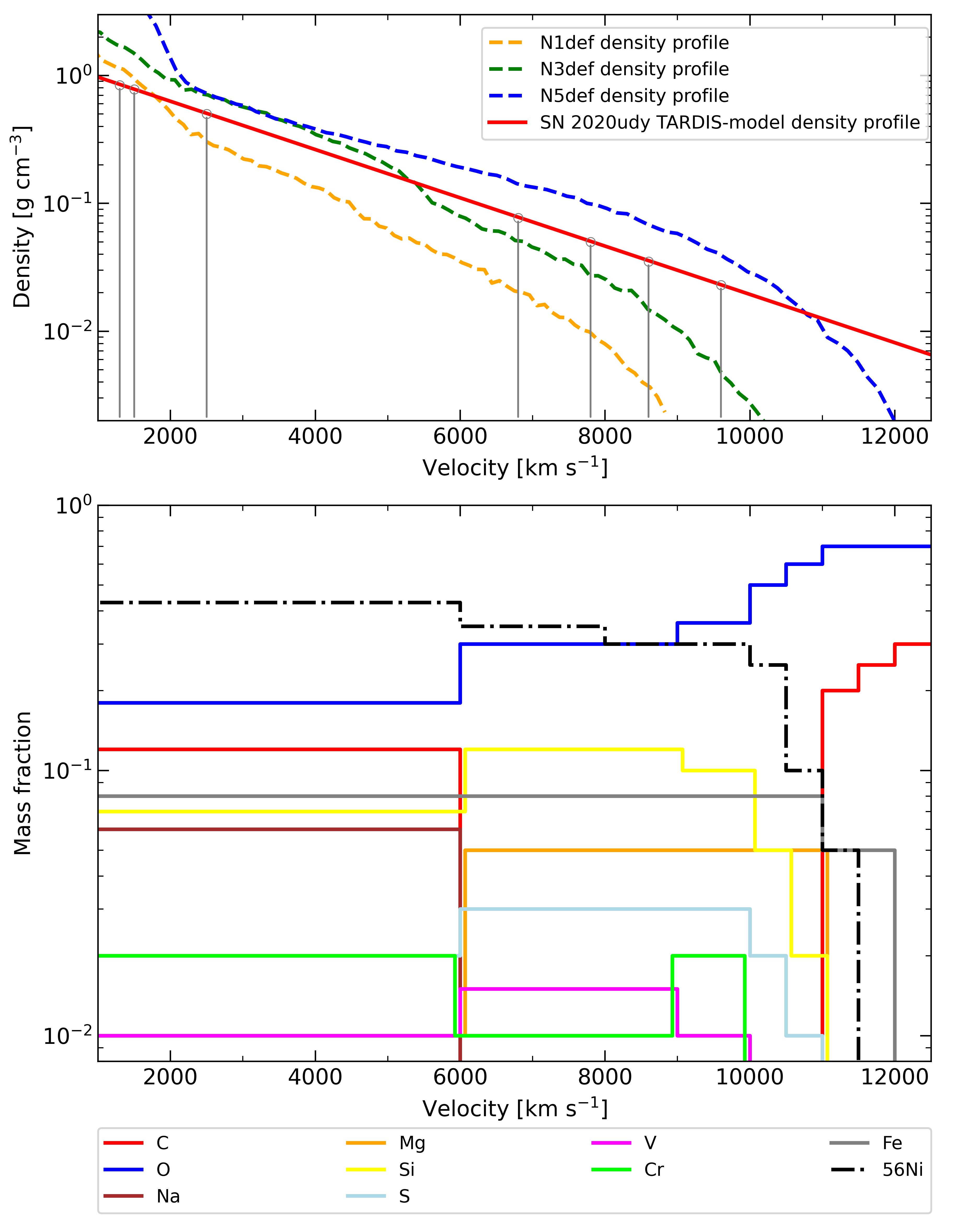}
	\end{center}
	\caption{Top panel: the best-fit TARDIS density profile (red) compared to the pure deflagration at $t_0 = 100$ s. The grey lines indicate the location of the photosphere for each of the analyzed epochs. Bottom panel: the best-fit chemical abundance structure from the fitting process for the spectral sequence of SN~2020udy. The profile of
the radioactive $^{56}$Ni shows the mass fractions at $t_\mathrm{exp}=100$ s.}
	\label{fig:SN_2020udy_abundance_density}
\end{figure}

The abundance tomography presented in this study is split into two parts: before $t_\mathrm{exp} = 30$ days from the date of explosion (hereafter referred to as early epochs), we follow the standard technique for synthesizing a spectral time series \citep{2023ApJ...953...93S}, while for the later epochs, a simplified method is adopted similar to that of \cite{2023ApJ...951...67C}. For both phases, the same time of the explosion ($T_\mathrm{exp}=2459116.3$), which is within the uncertainty range of the time derived from early light curve synthesis (see in Section \ref{explosion_epoch}), and density profile are used for the fit of any epochs. The latter one is chosen as a simple exponential function and constrained from the fitting as:

\begin{equation}
\rho (v,t_{\rm{exp}}) = \rho_0 \cdot \left(\frac{t_{\rm{exp}}}{t_0}\right)^{-3} \cdot \exp\left({-\frac{v}{v_0}}\right),
\end{equation}  

where $\rho_0 = 1.4$ g cm$^{-3}$ is the core density shortly after the time of the explosion ($T_\mathrm{exp}$) when the homologous expansion starts (here, chosen as $t_0=100$ s); while $v_0 = 2300$ km s$^{-1}$ is the exponential steepness of the density structure decreasing outwards. The chemical composition of ten elements (see in Figure \ref{fig:SN_2020udy_abundance_density}) is fit in radial shells with a velocity width of 500 km s$^{-1}$. Our initial assumption on each element's abundance was the constant fit of the abundance structure of the N5-def deflagration model \citep{2014MNRAS.438.1762F}. During the fitting process, we aimed for simplicity and modified the actual abundance structure of a shell only when the quality of the match between the synthetic and observed spectra is clearly improved with the change.
At later epochs ($t_\mathrm{exp} > 30$ days), the minor changes in abundance have little or no impact on the outcome of the radiative transfer, thus, we fixed the initially assumed constant values with minor modifications regarding the presence of lines of certain elements (see below). Note that despite several simplifications, the number of fitting parameters is way too high to fully cover the parameter space, thus, our 'best-fit' model can be considered as a feasible solution at its best.

\begin{figure*}
	\begin{center}
		\includegraphics[width=15cm]{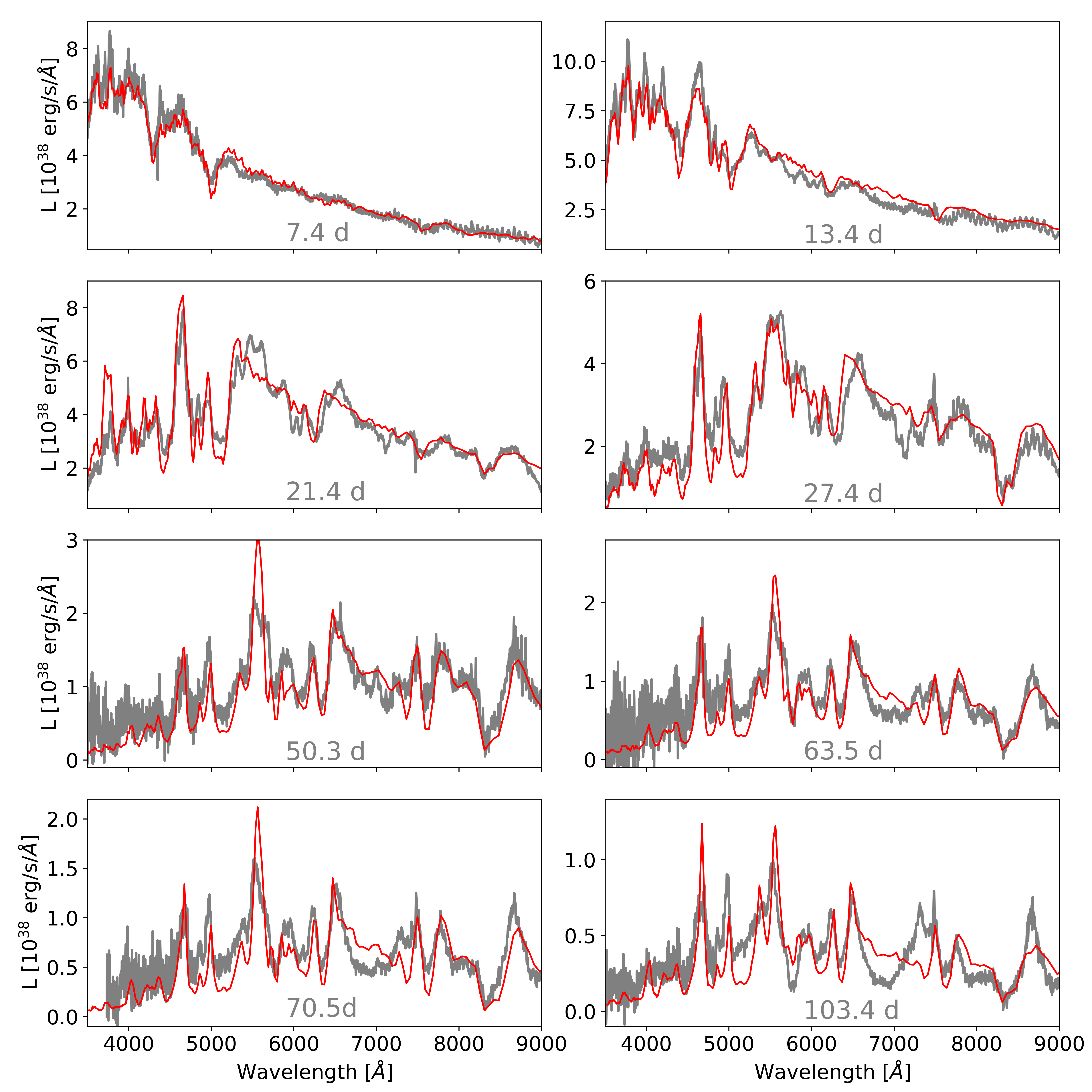}
	\end{center}
	\caption{Spectral synthesis of the evolution of SN~2020udy. The model spectra (in red) are produced with TARDIS within the framework of the abundance tomography analysis.}
	\label{fig:SN_2020udy_tardis}
\end{figure*}

The synthetic spectra are displayed in Figure \ref{fig:SN_2020udy_tardis}. The dominant spectral features of SN~2020udy are reproduced by our final model at every epoch. The absorption profiles of IMEs such as Si, S, Ca, etc. are nicely fitted at all phases. As an observable weakness of the fits, the P-Cygni features of IGEs after 20 days are overestimated in general (see between 4000 and 5500 \r{A} in Figure \ref{fig:SN_2020udy_tardis}). However, the fix of this issue involves either reduced Fe and $^{56}$Ni abundances or lower $v_0$ to decrease the outer densities; but both options would reduce the goodness of the fits at other epochs.

The best-fit density profile (upper panel of Figure \ref{fig:SN_2020udy_abundance_density}) shows the same steepness ($v_0 = 2300$ km\,s$^{-1}$) as that of other type Iax SNe, which were the subject of similar abundance tomography analysis \citep{Barna2018_mnras, 2021MNRAS.501.1078B}. The inferred $\rho(v)$ function scales the densities between those of the N3-def and N5-def models during the early epochs, which is consistent with Figure \ref{fig:SN_2020udy_deflagration}. The constrained abundance structure shows a strong stratification of both of the IGE and IME mass fractions that reproduce the fast evolution of spectral lines. As a further contradiction to the pure deflagration models, C is not allowed below 11\,000 km s$^{-1}$ in our model to prevent the appearance of extremely strong C II $\lambda$4618 and $\lambda$6578. The possible detection of such stratification in the outermost region of type Iax SNe is not unprecedented in the literature \citep[see e.g.][]{2013MNRAS.429.1156S,Barna2018_mnras}, but studies with other methodologies of spectral synthesis argued against chemical layering \citep{2022MNRAS.509.3580M}. Note that due to the high level of degeneracy of our fitting process, the results regarding the stratification are inconclusive. Optical spectroscopy at even earlier phases ($t_\mathrm{exp} < 4$ days) and/or UV spectral observations are required to resolve this issue.

The fits of the late-time epochs (below $\sim6000$ km s$^{-1}$) support the existence of a uniform chemical structure in the inner ejecta \citep{2014MNRAS.438.1762F}. Similar to \cite{2023ApJ...951...67C}, we set a high Na mass fraction to produce the P - Cygni feature at $\sim5800$ \r{A}. As a further modification, we increase the abundance of V and Cr to create the features observed around $\sim4000$ and $\sim5800$ \r{A}, respectively. While the overabundance of Na is well-justified due to the unambiguous identification, similar to that of the extremely late-time spectral synthesis of SN 2014dt, the presence of Cr and V with the 1-2\% mass fraction is not conclusive.

\section{Discussion}
\label{discussion}

The relatively complete photometric coverage of pre-maximum evolution of SN~2020udy allows us to constrain its rise time to $\sim$ 15 days in $B-$band, which is similar to SN 2005hk \citep{2007PASP..119..360P}. Our analysis of the pseudo bolometric light curve of SN~2020udy estimates the $^{56}$Ni mass of  0.08$\pm$0.01 M$_{\odot}$ and the ejecta mass of 1.39$\pm$0.09 M$_{\odot}$. Comparison of the pseudo bolometric light curves of SN~2020udy with different deflagration models shows that SN~2020udy lies between N3-def and N5-def models during the photospheric phase. Spectroscopic features and photospheric velocity evolution of SN~2020udy are similar to other bright type Iax SNe. The late nebular phase spectral features of SN~2020udy are broad and similar to SN 2012Z. To ascertain the most probable explosion scenario/progenitor system, we compare the observational properties of SN~2020udy with different proposed models for type Iax SNe. 

The PDD scenario starts with a slow deflagration which expands the white dwarf. Subsequently, infalling C-O layer triggers detonation \citep{1974Ap&SS..31..497I,1991A&A...245L..25K,1991A&A...246..383K,1993A&A...270..223K,1995ApJ...444..831H,1996ApJ...457..500H,2006ApJ...642L.157B,Baron_2012,2014MNRAS.441..532D}. The mass of $^{56}$Ni estimated in SN~2020udy is less than the mass of $^{56}$Ni produced in PDD scenario (0.12 to 0.66 M$_{\odot}$, \citealt{1995ApJ...444..831H}). The observed {\it (B-V)} color (0.33 mag) of SN~2020udy at maximum is slightly bluer than the {\it (B-V)} color (0.44 mag, \citealt{1995ApJ...444..831H}) expected in PDD54 model. 

In DDT models \citep{1991A&A...245L..25K,1991A&A...245..114K,1993A&A...270..223K,1995ApJ...444..831H,1996ApJ...457..500H,2002ApJ...568..791H,2013MNRAS.429.1156S,2013MNRAS.436..333S}, a deflagration flame is ignited by nuclear burning followed by expansion of the star. A considerable amount of the fuel burns into IMEs at lower ignition densities. Subsequently, a detonation commences which consumes most of the fuel. The mass of synthesized $^{56}$Ni in the case of SN~2020udy (0.08$\pm$0.01 M$_{\odot}$ ) is significantly less than the expected mass of synthesized $^{56}$Ni  produced in the DDT explosion (0.32 - 1.1 M$_{\odot}$, \citealt{2013MNRAS.436..333S}). Although the {\it (B-V)} color of SN~2020udy (0.33 mag) at maximum falls in the range of the DDT models (0.15 - 0.56 mag), other parameters such as the $B-$band absolute magnitude of SN~2020udy (M$_{B, max}$ = $-$17.41$\pm$0.34 mag) is not consistent with the DDT model prediction ($\sim-19.93$ to $-$18.16 mag). 

The three-dimensional pure deflagration of a white dwarf yields a range of synthesized $^{56}$Ni mass of 0.03--0.38\,M$_{\odot}$ \citep{2014MNRAS.438.1762F}. The rise time and peak absolute magnitudes provided by these models fall in the range of 7.6 to 14.4 days and $-$16.84 to $-$18.96 mag, respectively. Most of the observed parameters of SN~2020udy fit well within the range of parameters predicted by such a 3D pure deflagration process. Type Iax SNe are considered as a heterogeneous class, the progenitor and explosion scenario at the two extreme ends of luminosity distribution are shown to be different. However, bright members of this class (excluding outliers) behave in a similar manner and most of the observed properties of other bright type Iax SNe such as SNe 2020rea \citep{2022MNRAS.517.5617S}, 2020sck \citep{2022ApJ...925..217D}, etc., are consistent with the pure deflagration of a white dwarf in 3D \citep{2014MNRAS.438.1762F}. Based on the comparison of bolometric light curves and the density structure constrained from the abundance tomography analysis, SN~2020udy resembles a transition between the N3-def and N5-def pure deflagration models \citep{2014MNRAS.438.1762F}.

\section{Summary}
\label{summary}

The extensive photometric and spectroscopic follow-up of SN~2020udy reveals the following features: 

\begin{itemize}
    \item SN~2020udy is a bright member of the type Iax class with M$_{B, max}$ = $-$17.41$\pm$0.34 mag.

    \item The analytical modelling of the pseudo-bolometric light curve yields 0.08$\pm$0.01 M$_{\odot}$ of synthesized $^{56}$Ni with a well constrained rise time of $\sim$ 15 days.

    \item Spectroscopic features of SN~2020udy show similarity with the other bright members of the type Iax class. 
    
    \item Abundance tomography modeling of SN~2020udy shows photospheric velocity of $\sim$8000 km s$^{-1}$ at maximum. This fits into the general trend of the Iax class, as brighter objects expand with higher velocity.

    \item Comparison of proposed explosion models with the observational parameters shows that SN~2020udy is consistent with explosion models of pure deflagration of a white dwarf.
    
    \item While the earliest spectra, which sample the outermost layers of the ejecta, show some indications of chemical stratification, the post-maximum evolution of SN~2020udy is consistent with the predictions of the pure deflagration scenario. 

    \item Thus, SN~2020udy is an addition to the bright type Iax SNe population. The observed similarities of SN~2020udy with other bright Iax SNe indicate homogeneity within bright members of this class.
    
\end{itemize}


\section{acknowledgments}

We thank the anonymous referee for providing useful comments and suggestions towards improvement of the manuscript. We acknowledge Wiezmann Interactive Supernova data REPository http://wiserep.weizmann.ac.il (WISeREP) \citep{2012PASP..124..668Y}. This research has made use of the CfA Supernova Archive, which is funded in part by the National Science Foundation through grant AST 0907903. This research has made use of the NASA/IPAC Extragalactic Database (NED) which is operated by the Jet Propulsion Laboratory, California Institute of Technology, under contract with the National Aeronautics and Space Administration. RD acknowledges funds by ANID grant FONDECYT Postdoctorado Nº 3220449. The work of XW is supported by the National Natural Science Foundation of China (NSFC grants 12288102, 12033003, and 11633002), the Science Program (BS202002) and the Innovation Project (23CB061) of Beijing Academy of Science and Technology, and the Tencent Xplorer Prize. This work makes use of data obtained with the LCO Network. This research made use of TARDIS, a community-developed software package for spectral synthesis in supernovae \citep{kerzendorf_wolfgang_2018_1292315, kerzendorf_wolfgang_2019_2590539}. The development of TARDIS received support from the Google Summer of Code initiative and from ESA's Summer of Code in Space program. TARDIS makes extensive use of Astropy and PyNE. This work made use of the Heidelberg Supernova Model Archive (HESMA)\footnote{\url{https://hesma.h-its.org}}. DAH, GH, and CM were supported by NSF Grants AST-1313484 and AST-1911225. We acknowledge the usage of the HyperLeda database (http://leda.univ-lyon1.fr). The SDSS is managed by the Astrophysical Research Consortium for the Participating Institutions. The Participating Institutions are the American Museum of Natural History, Astrophysical Institute Potsdam, University of Basel, University of Cambridge, Case Western Reserve University, University of Chicago, Drexel University, Fermilab, the Institute for Advanced Study, the Japan Participation Group, Johns Hopkins University, the Joint Institute for Nuclear Astrophysics, the Kavli Institute for Particle Astrophysics and Cosmology, the Korean Scientist Group, the Chinese Academy of Sciences (LAMOST), Los Alamos National Laboratory, the Max-Planck-Institute for Astronomy (MPIA), the Max-Planck-Institute for Astrophysics (MPA), New Mexico State University, Ohio State University, University of Pittsburgh, University of Portsmouth, Princeton University, the United States Naval Observatory, and the University of Washington.

%

\vspace{5mm}






\bibliography{ms}{}
\bibliographystyle{aasjournal}



\end{document}